\documentclass[journal]{IEEEtran}
\pdfoutput=1 
\author{Khawaja Fahad Masood, Jun Tong, Jiangtao Xi, Jinhong Yuan and Yanguang Yu  
\thanks{K. F. Masood, J. Tong, J. Xi and Y. Yu are with School of Electrical, Computer and Telecommunications Engineering, University of Wollongong, Australia.}
\thanks{K. F. Masood is on leave from the Department of Electrical and Computer Engineering, COMSATS University of Science and Technology, Islamabad, Abbottabad Campus, KPK,
		Pakistan.}		
\thanks{J. Yuan is with the School of Electrical Engineering and Telecommunications, University of New South Wales, Australia. }
\thanks{This work has been accepted for publication in IEEE Transactions on Wireless Communications.  Copyright may be transferred without notice, after which this version may no longer be accessible. }}

\usepackage[dvips]{changebar}
\usepackage{soul}
\usepackage{amsmath}
 \usepackage{graphicx}
\usepackage{amssymb}
\usepackage{subcaption}
\usepackage{lineno}
\usepackage{makecell}
\usepackage{indentfirst}
\usepackage{bm}
\usepackage{color}
\usepackage{cite}
\usepackage{amsthm}
\usepackage{epstopdf}

\usepackage[noend]{algpseudocode}
\usepackage{amssymb}

 \usepackage{lipsum}
\usepackage{tablefootnote}
\usepackage{algorithmicx}
\usepackage{algpseudocode}

\usepackage[ruled,vlined]{algorithm2e}
\usepackage[font=footnotesize, skip=0pt]{caption}

\newcommand{\mb}{\mathbf}
\def\BibTeX{{\rm B\kern-.05em{\sc i\kern-.025em b}\kern-.08em
    T\kern-.1667em\lower.7ex\hbox{E}\kern-.125emX}}

\begin{document}
% 	\twocolumn[
% \begin{center}
% 	This work has been accepted for publication in IEEE Transactions on Wireless Communications.%  Copyright may be transferred without notice, after which this version may no longer be accessible.
% \end{center}]

\title{Inductive Matrix Completion and Root-MUSIC-Based Channel Estimation for Intelligent Reflecting Surface (IRS)-Aided Hybrid MIMO Systems}
         
\maketitle
\bibliographystyle{IEEEtran}
\begin{abstract}
This paper studies the  estimation of cascaded channels in passive intelligent reflective surface (IRS)-aided multiple-input multiple-output (MIMO) systems employing hybrid precoders and combiners. We propose a low-complexity solution that estimates the channel parameters progressively. The angles of departure (AoDs) and angles of arrival (AoAs) at the transmitter and receiver, respectively, are first estimated using inductive matrix completion (IMC) followed by root-MUSIC-based super-resolution spectrum estimation. Forward-backward spatial smoothing (FBSS) is applied to address the coherence issue. Using the estimated AoAs and AoDs,  the training precoders and combiners are then optimized and the angle differences between the  {AoAs and AoDs} at the IRS are estimated using the least squares (LS) method followed by FBSS and the root-MUSIC algorithm. Finally, the composite path gains of the cascaded channel are estimated using on-grid sparse recovery with a small-size dictionary. The simulation results suggest that the proposed estimator can achieve improved channel parameter estimation performance with lower complexity as compared to several recently reported alternatives, thanks to the exploitation of the knowledge of the array responses and low-rankness of the channel  using low-complexity algorithms at all the stages. 
\end{abstract}


\section{Introduction} 

Millimeter-wave (mmWave) and terahertz (THz) bands have gained significant interests for 5G and beyond thanks to their rich spectrum resources. Such high-frequency bands face higher penetration and path loss \cite{rappaport2013millimeter,xiao2017millimeter, ning2021prospective} compared to the sub-6 GHz bands. Massive multiple-input multiple-output (MIMO) systems can help compensate those losses. However, fully digital massive MIMO systems may lead to excessive power consumption. Hybrid transceiver architectures using switches and/or phase shifter networks can reduce the number of RF chains required, which can in turn alleviate the cost and power consumption of the hardware \cite{ning2021prospective, mendez2016hybrid}. In such systems, channel estimation is challenging due to the increased dimensionality and reduced baseband observations, especially when ultra-massive MIMO is considered for future-generation wireless communications \cite{yang20196g, zheng2022survey}.  

On the other hand, the propagation paths from the transmitter to the receiver at the high-frequency bands are typically fewer than those at the sub-6 GHz bands \cite{rappaport2013millimeter}. Consequently, mmWave and THz channels are more susceptible to the blockage of propagation paths, which affects the coverage and quality of service (QoS) of the system. Intelligent reflective surfaces (IRS) have emerged as a potential solution to address this challenge. IRS is usually constructed using passive reflective surfaces or  {meta-surfaces}. They can programmably alter the phase and/or amplitude of the incident signals at very low power consumption  \cite{zheng2022survey,di2020smart}. This offers a new degree of freedom  for more controllable wireless  environments and thus improve the   communication performance \cite{wu2019towards}. For example, in the absence of line-of-sight (LOS) paths between the transmitter (TX) and receiver (RX), IRS can provide alternative paths with strong gains to boost the coverage. Channel estimation is essential for designing the passive and active beamformers for IRS-aided MIMO systems. This is especially challenging as IRS is mostly passive.  
 
The channels of IRS-aided systems may be estimated in different manners, depending on whether the IRS possesses baseband signal processing capabilities. With separate channel estimation, the TX-to-IRS and IRS-to-RX subchannels are both estimated explicitly. This often requires a certain number of active elements to be deployed at the IRS, leading to semi-passive IRS, so that digital observations can be captured at the IRS. A relatively low training overhead may suffice but this costs increased complexity and power consumption of the IRS. Various schemes for separate channel estimation have been proposed, see \cite{alexandropoulos2020hardware,lin2021tensor,liu2020deep,jian2020modified} for examples.  

In this paper, we focus on fully passive IRS-aided MIMO systems with hybrid transceivers. This requires cascaded channel estimation based only on the observations at the hybrid  receiver. Due to the high dimensionality, the training overhead can be significantly increased using classical estimators such as the least squares (LS).  Tremendous efforts have been made to address this crucial challenge, such as \cite{he2019cascaded,hu2021two,wang2020channel,guan2021anchor,liu2021deep,kundu2021channel, wang2021channel,ye2022channel,wang2020compressed,ardah2021trice,liu2020matrix,he2020channel,he2021channel}, 
and a thorough survey can be found in \cite{zheng2022survey}. For example, \cite{he2019cascaded} exploits a bilinear matrix factorization model for the training data. Capitalizing on the sparsity and low-rankness of the factor matrices, they develop a two-stage estimator by using iterative sparse matrix factorization based on the bilinear generalized approximate
message passing (BiG-AMP) algorithm followed by low-rank matrix completion via Riemannian gradient. 
A two-timescale approach is studied in \cite{hu2021two} to reduce the training overhead by exploiting the quasi-static nature of the base station (BS)-to-IRS subchannel. This may require the BS operating in the full-duplex mode capable of self-interference mitigation.  In \cite{wang2020channel}, the redundancy in the multi-user cascaded channels is leveraged and a three-phase estimator using LS and linear minimum mean squared error (LMMSE) estimation is designed. Anchor-based solutions are investigated, e.g., in \cite{guan2021anchor}. 
There are also solutions exploiting machine learning (ML).  
For example, \cite{liu2021deep,kundu2021channel,wang2021channel} propose to denoise the (interpolated) LS estimates of the cascaded channel using neural networks. Their pilot overhead depends on the requirement of the LS estimation, which in turn depends on the dimensionality and the number of RF chains at the receiver.   
 A conditional generative adversarial network-based solution is also proposed in \cite{ye2022channel}.  
Many of the above solutions assume a single antenna at the users, and employ single-stage training, which may incur a substantial training overhead due to the lower beamforming gains achievable. Furthermore, they generally do not utilize the knowledge of the array responses at the transmitter, receiver or IRS, but aim to directly estimate the entries of the channel matrices.

For IRS-aided mmWave and THz MIMO systems, however, it may also be beneficial to exploit knowledge about the array responses for channel estimation. The associated parametric representations of the cascaded channel may then be employed to reduce the dimensionality of the signal processing problems. Instead of directly estimating the cascaded channel matrix or its factors, parameters such as the angles of arrival (AoAs), angles of departure (AoDs) and path gains can be estimated for acquiring the channel state information (CSI).  For example, 
in \cite{wang2020compressed}, the cascaded channel estimation is formulated as a sparse recovery problem, which is then solved using on-grid  compressive sensing (CS) algorithms including orthogonal matching pursuit (OMP) and generalized approximate message passing (GAMP). With this scheme, a multi-dimensional dictionary accounting for multiple path directions is used, which may have a big size that affects the computational complexity.  
 In \cite{ardah2021trice}, a similar sparse recovery formulation is developed while a two-stage approach is applied to solve the problem for achieving a lower complexity. First, the AoAs at the RX and AoDs at the TX are jointly estimated using 2D CS or super-resolution  spectrum estimation techniques (e.g., the beamspace ESPRIT). With such angle  information generated, the angles related to the IRS are next estimated using similar techniques. By decoupling the angle estimation in two stages, the complexity can be reduced but 2D CS or spectrum estimation is still required. 
 In \cite{liu2020matrix}, multiuser MIMO systems are considered and the channel estimation problem is formulated as a matrix-calibration based matrix factorization task. The channel sparsity over a predefined dictionary of the path angles and slow variations of the IRS-to-BS channel are exploited to develop a message passing-based algorithm for channel recovery. In \cite{he2020channel}, another two-stage estimator based on an iterative reweighted solution to a CS problem is proposed, which can address the grid errors to enhance performance. Note that \cite{wang2020compressed,ardah2021trice,he2020channel,liu2020matrix} all employ a single-stage training without exploiting prior  knowledge of the path directions. By contrast, in a recent work \cite{he2021channel}, two-stage training is introduced, where {Stage 1} aims to recover the AoAs and AoDs at the RX and TX, respectively, and {Stage 2} exploits those angle information to optimize the training scheme. Gridless spectrum estimation is adopted at both stages to recover the angles of interest. This scheme can benefit significantly from the gleaned angle information for achieving beamforming gains at a low feedback overhead. The parameters of the cascaded channel are estimated progressively, reducing the complexity as compared to approaches adopting joint parameter estimation. However, within each stage of the solution in \cite{he2021channel}, the computational complexity is high for large systems. This is due to the application of atomic norm minimization (ANM) to jointly exploit the low-rank property and the knowledge of the array responses for super-resolution estimation of the angles from compressed measurements, which requires semidefinite programming (SDP) for solving the multiple ANM problems involved.    

In this paper, we propose a low-complexity cascaded channel estimator for IRS-aided MIMO with hybrid transceivers. This estimator adopts the two-stage training/estimation framework originally proposed in \cite{he2021channel}: In Stage 1, fixed IRS phase shifts and varying transmitter precoders and receiver combiners are used for estimating the transmitter AoDs and receiver AoAs (referred to as the outer angles); In Stage 2, transmitter precoders and receiver combiners constructed based on the estimates of the outer angles and varying IRS phase shifts are used for estimating the IRS angle differences (referred to as the inner angles) and path gains. Despite using the same two-stage framework as \cite{he2021channel}, we propose novel multi-step treatments within each stage to exploit the low-rank property and the knowledge of the array responses using low-complexity algorithms in a progressive manner, so that super-resolution parameter estimation is achieved with low training overhead and low computational complexity. We also design new training schemes for each stage to enable the proposed estimation schemes. The proposed solution can yield a superior tradeoff among the estimation accuracy, training overhead, and computational complexity. Our contributions can be summarized as follows:

\begin{itemize} 

\item In Stage 1 of the proposed solution, we develop a structured training scheme (defined by the hybrid precoders and combiners) to enable an inductive matrix completion (IMC)-based approach for estimating an effective channel matrix, whose row and column subspaces embed the transmitter AoDs and receiver AoAs, respectively. The innovative integration of IMC and root-MUSIC (aided by forward-backward spatial smoothing (FBSS)) effectively exploits their respective strengths such as low training overhead and super-resolution parameter estimation, while addressing their respective challenges such as IMC's lack of exploitation of array responses and root-MUSIC's difficulty in handling compressed measurements. Consequently, we achieve super-resolution estimation of the outer angles with significantly reduced complexity as compared to the approach based on ANM and SDP.

\item In Stage 2, we develop a subarray-based training scheme to enable low-overhead estimation of the inner angles. The estimation scheme based on least squares (LS), FBSS and root-MUSIC achieves super-resolution estimation of the inner angles with a low complexity. The joint estimation of the angles is more robust in scenarios with imperfect estimates of the outer angles and finite-size antenna arrays and can thus improve the accuracy as compared to the separate estimation of the individual angles in \cite{he2021channel}. Furthermore, the proposed scheme exhibits much lower complexities as compared to ANM schemes  implemented with multiple usage of SDP. We further develop a compressive sensing (CS)-based approach for automatically associating the separately estimated inner and outer angles, which also yields the estimates of the composite path gains. 

\item We extend the proposed solution for the case with uniform planner arrays (UPAs) at the IRS. An L-shaped subarray sampling design is suggested for achieving high-resolution estimates of the azimuth and elevation angle differences at the IRS with a low training overhead and low computation complexity.

\end{itemize} 
 Simulation studies are performed to compare the proposed estimator with several recently proposed estimators.
	It is shown that high-accuracy estimation of the channel parameters can be achieved by the proposed solution, which may 
	yield better performance when there are multiple paths in the channel and the numbers of antennas and training overhead are limited.
	We also carry out a detailed analysis of 
	the computational complexity. 

The rest of the paper is organized as follows. Section II introduces the system  model.  Section III presents the proposed solution. The simulation results are discussed in Section IV and  the conclusions are drawn in Section V.

\begin{figure}[h]
	\twocolumn[\centering{	\includegraphics[width=5.5in ]{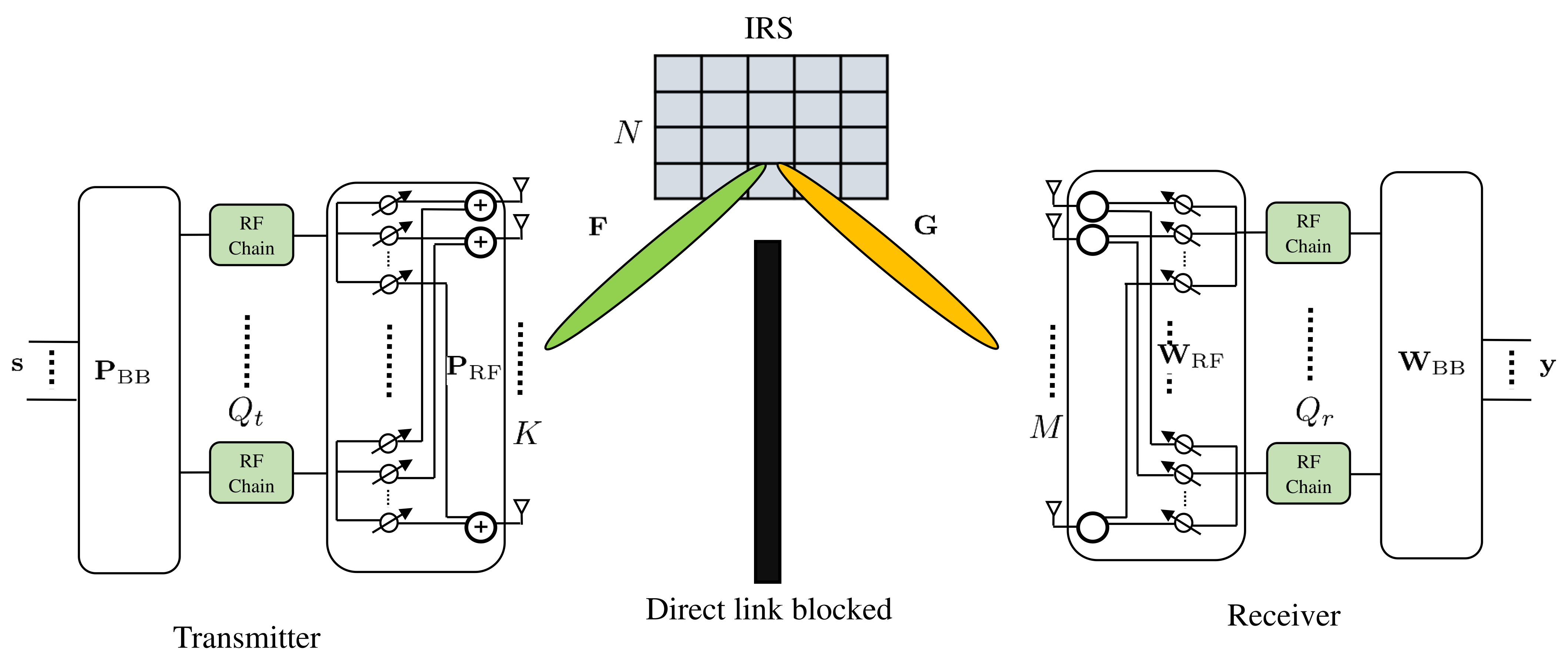} 
		\caption{System model.}\label{figsm}}]
\end{figure}
 {\bf Notation:}   
Throughout the paper, boldface capital letters and boldface lower-case letters such as $\mb X$ and $\mb x$ denote  matrices and vectors, respectively; $(\cdot)^T$, $(\cdot)^*$, $(\cdot)^H$, $(\cdot)^{-1}$ and $(\cdot)^\dagger$ represent matrix transpose, conjugate, conjugate transpose,  inverse and pseudo inverse, respectively; $\mathrm{vec}(\cdot)$, $\mathrm{Mat}(\cdot)$ and $\mathrm{diag}(\cdot)$ denote the vectorization, matricization and selection of diagonal entries of a matrix, respectively; and $\otimes$, $\diamond$ and $ \|\cdot\|_{F}$ denote the Kronecker product, Khatri-Rao product and Frobenius norm, respectively.

\section{System Model}

As illustrated in Fig. \ref{figsm}, we consider a point-to-point MIMO system equipped with hybrid transceivers\footnote{The proposed solutions can also be directly applied to fully digital systems which can be regarded as a special case of the hybrid systems when each antenna is equipped with a dedicated RF chain.} aided by a fully passive IRS. The transmitter, receiver, and IRS employ ULAs with $K$, $M$ and $N$ array elements, respectively\footnote{The techniques proposed by this work can be extended to multiple-user systems with ULA and UPA applied at the transmitter, receiver and IRS. In Section \ref{ExtUPAIRS} we will discuss the treatment for UPA at the IRS.}. There are $Q_t\le K$ and $Q_r\le M$ RF chains at the transmitter and receiver, respectively. Fully connected phase shifter networks are assumed for the transmitter and receiver but the techniques can also be easily extended to other hybrid transceivers \cite{masood2021low}.   The direct channel between the transmitter and receiver is assumed to be blocked for simplicity, but there exist many viable solutions \cite{lee2016channel,hu2018matrix,vlachos2018massive,masood2021low, liu2022lowrank}  to estimate it.

The channel between the transmitter and the IRS is geometrically modeled as \cite{he2021channel,ardah2021trice,zhou2022channel,schroeder2022two},
\begin{equation} 
\label{Fmat}
 \begin{aligned}
\mb F&=  {\sqrt{\frac{K N}{L_{F}}}} 
\sum_{l=1}^{L_{F}} \gamma_{F,l}  \mb a_I (\theta_{I,l})\mb a_T^H(\phi_{T,l})  
 \\& =\mb A_I (\bm \theta_I)\mb \Gamma_{F}\mb A_T^H (\bm \phi_{T})  \in \mathbb{C}^{N\times K},
\end{aligned}
\end{equation} 
where $\theta_{I,l}$, $ \phi_{T,l}$ and $\gamma_{F,l}$  represent the AoA at the IRS, the AoD at the transmitter and the complex path gain for the $l$-th path, respectively, and $L_F$ denotes the total number of paths between the  transmitter and IRS. Furthermore, $\mb A_I (\bm \theta_I)$, $\mb A_T(\bm \phi_T)$ and $\mb \Gamma_F$ denote the IRS array response matrix, transmitter array response matrix, and diagonal path gain matrix, respectively. Similarly, the channel between the IRS and the receiver is modeled as  
\begin{equation} 
\label{Gmat}
 \begin{aligned}
\mb G&=  { \sqrt{\frac{N M}{L_G}}} 
\sum_{l=1}^{L_G} \gamma_{G,l} \mb a_R (\theta_{R,l})\mb a_I^H (\phi_{I,l}) \\  &=\mb A_R (\bm \theta_{R})\mb \Gamma_{G}\mb A_I^H (\bm \phi_I) \in \mathbb{C}^{M\times N},
\end{aligned}
\end{equation} 
where  {notation similar to those in (\ref{Fmat}) is used.}
 {In the above, the array response vector for a ULA with $N$ elements can be written as}
\begin{align}
\mb a(\alpha)=\frac{1}{\sqrt{N}}[1, {e}^{j\frac{2\pi}{\lambda_c}d\cos(\alpha)},\dots, {e}^{j(N-1)\frac{2\pi}{\lambda_c}d\cos(\alpha)}]^T,
\end{align}
where $\lambda_c$ is the wavelength, $d=\lambda_{c}/2$ is the inter-element spacing, and $\alpha$ is the steering angle. We focus on mmWave and THz bands where the number of paths $L_F$ and $L_G$ are typically small.

  The  effective channel between the transmitter and receiver via IRS is given as 
\begin{align}
\bm {\mathcal  H} =\mb G\mb{\Omega}\mb F  \in \mathbb{C}^{M \times K},
\end{align}
where $\mb \Omega=\mbox{diag}(\boldsymbol{\omega})$ and $\boldsymbol{\omega}$ contains the phase shifts for all the IRS elements: 
\begin{align*}
\boldsymbol{\omega}=[\beta_1 {e}^{j\zeta_1}, \beta_2 {e}^{j\zeta_2}, \dots, \beta_{N} {e}^{j\zeta_{N}} ]^T\in \mathbb{C}^{N\times 1}, 
\end{align*} 
where $\beta_i$ and  $\zeta_i$ denote the reflection coefficient and the phase shift for the $i$-th IRS  element, respectively. Setting  $\beta_i=1$ or $0$ suggests that the $i$-th element is turned on or off, respectively.

By using the identity $\mathrm{vec}(\mb A \mbox{diag}(\mb b) \mb C)= (\mb C^T\diamond\mb A) \mb b$,  the  effective channel can be rewritten as  
\begin{equation} \label{Cascaded}
\begin{aligned}
  \mathrm{vec} (\bm {\mathcal  H} )&=\mbox{vec}(\mb G\mb{\Omega}\mb F)  = (\mb F^T\diamond\mb G) \boldsymbol{\omega} =\mb H \boldsymbol{\omega}, 
\end{aligned}
\end{equation} 
where $\mb H\triangleq\mb F^T\diamond\mb G \in \mathbb{C}^{MK\times N}$ is the cascaded channel. During data transmissions, the precoder, combiner and 
IRS phase shifts need to be optimized according to $\mb H$. However, it is challenging to estimate $\mb H$ due to its high dimensionality and limited observations at the hybrid receivers. In order to address this problem, we propose below a solution with low training overhead and low computational complexity.

\begin{figure} [h]
	\twocolumn[\centering{ \includegraphics[width=5in ]{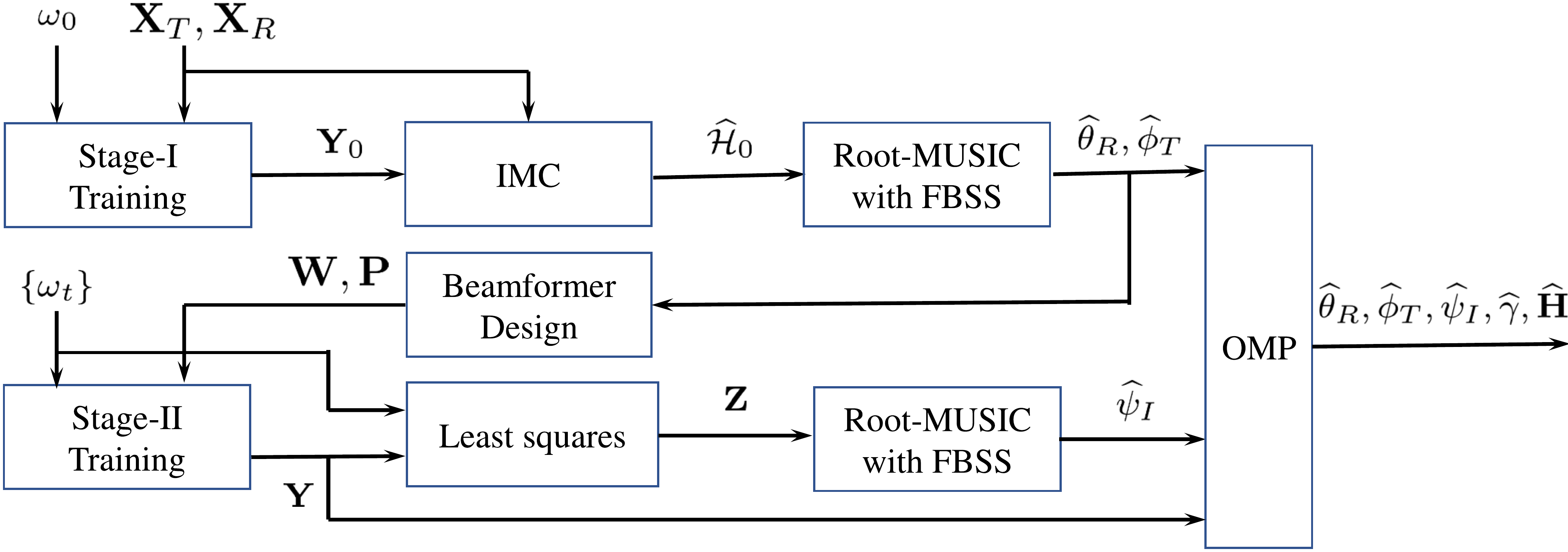} 
		\caption{Flowchart of the proposed scheme for estimating the cascaded channel of IRS-aided MIMO systems.} \label{flowchart}}]
\end{figure}

\section{ The Proposed Channel Estimator}

\subsection{Parametric Representation of the Cascaded Channel} 

By using the  identities  $(\mb A\mb B) \diamond (\mb C \mb D)=(\mb A \otimes \mb C) (\mb B \diamond \mb D) $ and $(\mb A \otimes \mb B) (\mb C \otimes \mb D)=(\mb A\mb C) \otimes (\mb B \mb D) $, the cascaded channel 
from (\ref{Cascaded}) can be modeled as 
\begin{equation}  
\label{HParameterized}
\begin{aligned}
\mb H 
&=(\mb A_I (\bm\theta_{I})\mb \Gamma_{F} \mb A_T^H(\bm\phi_{T}) )^T\diamond  \left(\mb A_R (\bm \theta_{R})\mb \Gamma_{G}\mb A_I^H(\bm \phi_{I}) \right)  \\  
&=( \mb A_T^*(\bm\phi_{T}) \mb \Gamma_{F} \mb A_I^T (\bm \theta_{I}) )\diamond  (\mb A_R(\bm \theta_{R})\mb \Gamma_{G}\mb A_I^H(\bm \phi_{I})) \\  
& = \left( (  \mb A_T ^*(\bm\phi_{T}) \mb \Gamma_{F} ) \otimes (\mb A_R (\bm \theta_{R})\mb \Gamma_{G}) \right)  \left( \mb A_I^T(\bm\theta_{I})\diamond \mb A_I^H(\bm\phi_{I}) \right)  \\ 
&=  {\mb A_{TR}(\bm \phi_T, \bm \theta_R) }   {\mb \Gamma}    {\mb  A_I^H(\bm \psi_I)},
\end{aligned}
\end{equation}  
where 
\begin{equation}
\label{Gamma_Matrix}
\bm \Gamma =   \mb \Gamma_{F}\otimes \mb \Gamma_{G} = \mbox{diag}(\bm \gamma),  
\end{equation}
with $\bm \gamma \in \mathbb{C}^{ L_FL_G \times 1}$ containing the composite of the channel path gains of  $\mb G$ and $\mb F$, 
\begin{equation}
\mb A_{TR}(\bm \phi_T, \bm \theta_R)  =   \mb A_T^*(\bm\phi_{T})  \otimes \mb A_R (\bm \theta_{R}), 
\end{equation}
and
\begin{align} \nonumber
\mb A_I(\bm \psi_I) &=\left(\mb A_I^H(\bm \theta_{I}) \diamond \mb A_I^T (\bm \phi_{I})\right)^T  \\& =[\mb a_I (\psi_{I,{1,1}}), \dots, \mb a_I (\psi_{I,{i,j}}),\dots, \mb a_I (\psi_{I,{L_F,L_G}})]
\end{align} 
 with 
\begin{align}
\label{anglediff}
\psi_{I,{i,j}}=\cos^{-1} (\cos(\phi_{I,j})-\cos(\theta_{I,i}))
\end{align} 
being the effective angle difference between the $i$-th AoA and $j$-th AoD at the IRS.    
From (\ref{HParameterized}), though the cascaded channel $\mb H$ may have a very large dimension, it can be parameterized with a small number of path directions and gains. This can be exploited to reduce the training overhead and computational complexity. However, joint estimation of these parameters may involve high computational complexities. We therefore propose a low-complexity, multi-stage solution below.

As shown in Fig. \ref{flowchart}, the proposed solution estimates the outer angles $\boldsymbol \phi_{T}$ and $\boldsymbol \theta_{R} $ in the first stage using varying hybrid precoder/combiners at the transmitter and receiver but fixed phase shifts at the IRS. IMC and spectrum estimation using the root-MUSIC algorithm are applied to estimate the angles. In the second stage, the IRS angles $\bm \psi_I$  are estimated using fixed hybrid precoders/combiners constructed using the estimated outer angles and varying IRS phase shifts.  Similarly, the root-MUSIC algorithm is applied at this stage. Finally the estimated angles are associated by solving a small-size on-grid CS problem using OMP, which also yields the composite path gains. The proposed solution has similarities to \cite{he2021channel} in terms of two-stage training and progressive estimation of the channel parameter. However, different training and estimation schemes are deployed, which can improve the complexity-performance tradeoff. 

It should be noted that each of the techniques employed such as the IMC, root-MUSIC and OMP may be used as a standalone solution for estimating the cascaded channel in IRS-aided hybrid MIMO systems. However, they face significant challenges: IMC lacks the exploitation of the knowledge of the array responses and does not provide estimates of path angles, root-MUSIC requires abundant uncompressed training samples, while OMP relies on a discrete dictionary. Our proposed solution integrates these techniques in a muti-step manner to exploit their respective strengths: IMC is capable of reconstructing large low-rank matrices from a small number of measurements. This can provide samples for root-MUSIC to exploit the array response for super-resolution parameter estimation, based on which a dictionary can be constructed for OMP to discover the sparsity pattern. Consequently, superior overall performance can be achieved with low training overhead and low computation complexity.

\subsection{Stage 1: Estimation of Outer Angles}
\label{stageoneproposed}

\subsubsection{Training}
In order to estimate the outer angles $(\boldsymbol{\theta} _{R}, \boldsymbol{\phi}_{T})$, the IRS phase shifts are randomly chosen as $\boldsymbol{\Omega}_0 = \mathrm{diag}(\bm \omega_0)$ from the feasible set and remain unchanged\footnote{The IRS phase shifts should be chosen such that the possibility of missing certain IRS AoAs or AoDs is low. When there are a large number of IRS elements, this can be achieved by randomly generating the IRS phase shifts.}. This gives the effective channel 
\begin{equation} 
\label{mathcalH0}
\bm {\mathcal  H} _0 =\mb G\mb{\Omega}_0\mb F \in \mathbb{C}^{M \times K}.
\end{equation}   
For mmWave and THz channels, $\mb G$ and $\mb F$ are generally low-rank due to the sparsity in the angular domain.  Therefore, $\bm {\mathcal H}_0$ is also low-rank with rank no higher than $\min(\mbox{rank} (\mb F), \mbox{rank}(\mb G))$. Furthermore, $\bm {\mathcal H}_0 $ can be modeled as  
\begin{align}
\label{Ho}
\bm {\mathcal H}_0  = \mb A_R(\bm \theta_{R})\mb \Gamma_{G}\mb A_I^H (\bm  \phi_{I}) \mb \Omega_0 \mb A_I (\bm  \theta_{I}) \mb \Gamma_{F} \mb A_T^H (\bm  \phi_{T}).   
\end{align}
From (\ref{Ho}), if $\bm {\mathcal H}_0$ is known, then subspace methods such as the root-MUSIC may be used to estimate the angles $(\bm \theta_R, \bm \phi_T)$, similar to the treatments in \cite{masood2021low}. In order to reduce the training overhead for estimating $\bm {\mathcal H}_0$ using the hybrid receiver, we propose to estimate $\bm {\mathcal H}_0$ using low-rank matrix recovery methods. We adopt the IMC scheme \cite{hu2018matrix} for its low complexity and high performance. 

We now describe the training scheme. Assume a training length of $S$ channel uses. 
During the $s$-th channel use, the transmitter sends a single pilot symbol $x_s$. The receiver observes $Q_r$ symbols through its RF chains:  
\begin{align}
\mb y_{s}=\mb W_{s}^H \bm {\mathcal  H} _0 \mb p_{s} x_{s}+\mb W_{s}^H\mb n_{s} \in \mathbb{C}^{Q_r \times 1},
\end{align} 
where $\mb p_{s}=\mb P_{\mathrm{RF}, {s}}\mb p_{\mathrm{BB}, s} \in \mathbb{C}^{K\times 1}$ is the hybrid precoder with the RF precoder $\mb P_{\mathrm{RF}, {s}}\in \mathbb{C}^{K \times Q_t}$ and baseband precoder $\mb p_{\mathrm{BB}, {s}}\in \mathbb{C}^{Q_t \times 1}$. Similarly, the hybrid combiner $\mb W_{s}=\mb W_{\mathrm{RF}, {s}}\mb W_{\mathrm{BB}, {s}} \in \mathbb{C}^{M \times Q_r}$ with $\mb W_{\mathrm{RF}, s}\in \mathbb{C}^{M\times Q_r}$ as the RF combiner and $\mb W_{\mathrm{BB}, {s}}\in \mathbb{C}^{Q_r\times Q_r }$ the basedband combiner.  
Without loss of generality, we assume $x_{s}=1$, $||\mb p_s||_F^2=1$ and $||\mb W_s||_F^2=Q_r $. The noise $\mb n_{s} \in \mathbb{C}^{M \times 1} \sim \mathcal{CN}(\mb 0, \sigma_n^2 \mb I)$, where $\sigma_n^2$ is the average noise power. 
The received signal after $S$ training steps   is given as
\begin{align}
\mb y_0 = [\mb y_{1}^T, \mb y_{2}^T,\dots, \mb y_{S}^T]^T \in \mathbb{C}^{S Q_r \times 1}.
\end{align}
 We aim to estimate $\bm {\mathcal H}_0$ from $\mb y_0$. For fully connected hybrid transceivers, one approach for directly estimating $\bm {\mathcal H}_0$ is to employ the specially tailored hybrid precoder/combiner design of \cite{hu2018matrix} such that   $\mb y_0$ consists of only noisy entries of $\bm {\mathcal  H} _0$ and matrix completion (MC) can be used to directly find $\bm {\mathcal  H} _0$. Another approach is to use low-rank matrix sensing (LRMS) similar to \cite{hu2019channel} for general precoders and combiners. However, the former approach requires extra training overhead while the latter requires higher computational complexity.

In order to achieve good performance while avoiding the limitations of the above-mentioned direct estimation of $\bm {\mathcal  H} _0$, we apply the IMC approach that first estimates the following transformed matrix   
\begin{align}
\label{IMCprob}
{\bm {\mathcal C}}_0  = \mb X_R^H \bm {\mathcal H}_0 \mb X_T \in \mathbb{C}^{M \times K},  
\end{align}
and then recover $\bm {\mathcal H}_0$ as ${\bm {\mathcal H}}_0 =({{\mathbf X}^{H}_{R}})^{-1}  {{\bm {\mathcal C}}_0 } ({\mathbf X}_{T})^{-1}$ 
where $\mb X_T$ and $\mb X_R$ are the feature matrices. 
The hybrid precoders and combiners are designed in a way such that 
entries of ${\bm {\mathcal C}}_0$ are observed from the above training process. This is implemented by selecting $\mb W_{s}$ and $\mb p_{s}$ from the columns of $\mb X_R \in \mathbb{C}^{M\times M}$ and $\mb X_T \in \mathbb{C}^{K \times K}$, respectively, following the uniform spatial sampling \cite{weng2012low}.      
Discussion on the coherence properties of $\mb X_R$ and $\mb X_T$ can be found in \cite[ Section D]{{hu2018matrix}} and the implementation of the hybrid 
precoders/combiners follows \cite[ Section II-B]{{masood2021low}}.

\subsubsection{Estimation of outer angles} Rewrite the received signal by $\mb Y_0=P_{\Omega }(\widetilde{\bm {\mathcal C}}_0)$, where $P_{\Omega }$ denotes the sampling operator with a sampling pattern $\Omega$ and $\widetilde{\bm {\mathcal C}}_0$ is a noisy version of ${\bm {\mathcal C}}_0$, i.e. $\widetilde{\bm {\mathcal C}}_0={\bm {\mathcal C}}_0+\mb X_R^H\mb N$, where $\mb N \in \mathbb{C}^{ M \times K}$ is the noise matrix. 
We can now estimate $(\boldsymbol \phi_{T}, \boldsymbol \theta_{R})$ from $\mb Y_0=P_{\Omega }(\widetilde{\bm {\mathcal C}}_0 )$ using IMC followed by spectrum estimation. The transformed matrix ${\bm {\mathcal C}}_0  =\mb X^H_R \bm {\mathcal H}_0\mb X_T$ is low-rank as $\bm {\mathcal H}_0$ is low-rank. Therefore, ${\bm {\mathcal C}}_0 $ can be estimated first by solving the following low-rank matrix recovery problem
 \begin{equation} 
\min _{{\bm {\mathcal C}}_0  } \mathrm{rank}({\bm {\mathcal C}}_0 ), \quad \mathrm {s.t.} \quad \|P_{\Omega }({\bm {\mathcal C}}_0) -\mb Y_0\|^{2}_{F}\leq \delta ^{2},
\end{equation} 
{where $\delta ^{2}$ is a tolerance to account for the noise.} Nuclear norm regularization is applied to reformulate the NP-hard problem above to estimate ${\bm {\mathcal C}}_0$ as
\begin{equation}  \label{CstarHat}
{\widehat{\bm {\mathcal C}}_0 } \triangleq 
\arg\min _{{\bm {\mathcal C}}_0} \quad \frac {1}{2}\|P_{\Omega }(\bm {\mathcal C}_0) -\mb Y_0\|_{F}^{2} + \mu \|{\bm {\mathcal C}}_0 \|_{\ast },\end{equation}
where $\mu>0$ is a regularization parameter and  {$ \|\cdot\|_{\ast}$ represents the nuclear norm}. By using a Frobenius norm characterization of the nuclear norm, we can let $\bm {\mathcal C}_0 \triangleq \mb U\mb V^H$ and recover $\bm {\mathcal C}_0$ by solving  
	\begin{align}
	\label{Main_eq} 
	\min_{{\mb U},{\mb V}}  \frac{1}{2} \| P_\Omega(\mb U\mb V^H)- \mb Y_0 \|^2_F +\frac{1}{2}\mu(\|\mb U\|^2_F+\|\mb V\|^2_F). 
	\end{align}
This is a regularized least squares problem if the sizes of $\mb U$ and $\mb V$ are fixed according to the rank of $\bm {\mathcal C}_0$.  {However, this rank is unknown in practice}. 
Various low-rank matrix recovery algorithms may be used to solve (\ref{Main_eq}) approximately. 
We adopt the generalized conditional gradient-based alternate minimization algorithm (GCG-ALTMIN)  \cite{hu2018matrix}  that progressively increases the sizes of $\mb U$ and $\mb V$  by using the top singular vectors of a residual error matrix and alternately refining $\mb U$ and $\mb V$ using local minimization. This algorithm avoids the computation of full singular value decompositions (SVDs) and benefits from the fast convergence by combining spectral initialization and alternate minimization. Consequently, the GCG-ALTMIN algorithm can recover ${{\bm {\mathcal C}}_0 }$ with lower complexity than many other alternative algorithms.  

Once the transformed matrix $\widehat{\bm {\mathcal C}}_0 $ is estimated, the {low-rank} channel matrix  can be estimated as 
\begin{equation} 
\label{Hl}
\widehat{\bm {\mathcal H}}_0 =({{\mathbf X}^{H}_{R}})^{-1}  {\widehat{\bm {\mathcal C}}_0 } ({\mathbf X}_{T})^{-1}.
\end{equation}
We can then estimate the outer angles $\bm \phi_{T}$ and $\bm \theta_{R}$.  
Following \cite{masood2021low}, we estimate them separately to reduce the computational cost. In this work, we apply the root-MUSIC algorithm \cite{rao1989performance, friedlander1993root, vesa2010direction}, 
which avoids peak search and offers high-resolution estimates of the angles with low complexity.
Note that the channel estimate $\widehat{\bm {\mathcal H}}_0$ in (\ref{Hl}) can be modeled as 
\begin{equation} 
 \label{HhatL}  
\begin{aligned}
\widehat{\bm {\mathcal H}}_0&=\bm {\mathcal  H}_0 +\mathbf E  =\mb A_R (\boldsymbol \theta_R) \mb G_0 \mb A^H_T(\bm \phi_T) +\mathbf E, 
\end{aligned}
\end{equation} 
where $\mathbf E$ represents the estimation error and $\mb G_0 \triangleq \mb \Gamma_{G}\mb A_I^H (\bm \phi_{I})\mb{\Omega_0}\mb A_I (\bm \theta_{I}) \mb \Gamma_{F}$. 
It is clear that the row and column subspaces of $ {\bm {\mathcal H}}_0$ are spanned by the receiver and transmitter steering vectors. This can be utilized to estimate the AoAs and AoDs separately by using subspace methods. 

We first estimate the AoAs $\bm \theta_R$ at the receiver. The estimation of the AoDs $\bm \phi_T$ at the transmitter is similar.   We model the columns of $\widehat{\bm {\mathcal H}}_0$ as samples of the received signal of an $M\times 1$ ULA as
\begin{equation}
\label{AoAEst}
\mb x_k = \mathbf A_R(\bm \theta_R) \bm \lambda_k + \mb e_k,  k=1,2,\dots, K, 
\end{equation} 
where $\bm \lambda_k$ is the $k$-th column of $\mb G_0 \mb A^H_T(\bm \phi_T)$ that serves as the ``source" for generating the observation $\mb x_k$ at the receiver array. The root-MUSIC algorithm can be applied using the signal and noise subspaces  {estimated} from the sample covariance matrix (SCM) of $\mb x_k$: 
\begin{equation} 
\begin{aligned}
\label{RAhatL}
\widehat{\mathbf R}_{\bm \theta_R} &=\frac{1}{K} \sum_{k=1}^K \mb x_k \mb x_k^H= \frac{1}{K} \widehat{\bm {\mathcal H}}_0	\widehat{\bm {\mathcal H}}_0^H  	\\&= \mathbf A_R (\bm \theta_{R})\bm \Delta \mathbf A_R^H(\bm \theta_{R}) + \bm \Sigma,     
\end{aligned}
\end{equation} 
where the ``source" covariance matrix  is given by  
\begin{equation}
 \label{GammaA}
\bm \Delta  \triangleq \frac{1}{K}  \mathbf G_0 \mathbf A_T^H(\bm \phi_{T})\mathbf A_T(\bm \phi_{T})\mathbf G_0^H, 
\end{equation} 
and the error of the covariance matrix estimation is 
	\begin{equation} \label{GammaA}
	\bm \Sigma \triangleq  \frac{1}{K} \left( \bm {\mathcal H}_0 \mb E^H + \mb E \bm {\mathcal H}_0^H + \mb E \mb E^H \right).  
	\end{equation}
Eigenvalue decomposition of $\widehat{\mathbf R}_{\bm \theta_R}$ can be used to find the  signal and noise subspaces required by root-MUSIC.

In general, the ``source" covariance matrix $\bm \Delta $ in (\ref{RAhatL}) is non-diagonal. This suggests that the ``source" signals in the model of (\ref{AoAEst}) are correlated. It is known that with correlated sources, standard subspace methods based on the SCM may perform poorly. We thus adopt the FBSS  \cite{shan1985spatial,pillai1989forward,van2004optimum} here to improve angle estimation. 

In order to estimate $L$ angles using the $M$-element array (\ref{AoAEst}), the FBSS constructs $U=L+1$ forward and backward uniform sub-arrays, each with $\mathcal{S}=M-L$ elements. Neighboring subarrays differ by only one element. Consider a reference ULA subarray with $\mathcal{S}$ antennas. Its array response matrix can be written as 
	\begin{equation}
	\widetilde{\mb A}_R ({\bm\theta}_R)= 
	[
	\widetilde{\mb a}_R (\theta_{R,1}),   \widetilde{\mb a}_R (\theta_{R,2}), \dots,  \widetilde{\mb a}_R (\theta_{R,L})
	] \in \mathbb{C}^{\mathcal{S} \times L},
	\end{equation}
	with array response for the $l$-th ``source" given as
	\begin{equation}
	\widetilde{\mb a}_R (\theta_{R,l})= \frac{1}{\sqrt{M}}[1,   e^{j \frac{2\pi}{\lambda_c}d\cos(\theta_{R,l})},  \dots,  {e}^{j ( \mathcal{S}-1)\frac{2\pi}{\lambda_c}d\cos(\theta_{R, l})}
	]^T.
	\end{equation}
	Let $\mb D$ 
	be a diagonal matrix with entries
	\begin{equation}
	\mb D=\mbox{diag}\{   {e}^{j\frac{2\pi}{\lambda_c}d\cos(\theta_{R,1})},  \mbox{e}^{j\frac{2\pi}{\lambda_c}d\cos(\theta_{R,2})},  \dots,   {e}^{j  \frac{2\pi}{\lambda_c}d\cos(\theta_{R,L})}\}. 
	\end{equation}
	Then the received signals of the $u$-th forward subarray, whose array response vectors are shifted versions of those of the reference subarray, can be written as 
	\begin{equation} 
	\begin{aligned}
	\mb x_{k, u}^f & \triangleq [
	x_{k,u},  x_{k,u+1},  \dots,  x_{k, u+\mathcal{S}-1}]^T  \\&= \widetilde{\mb A} _R (\bm\theta_R)\mb D^{(u-1)}  {\bm \lambda}_{k } + \mb e_{k,u}^f,
	\end{aligned}
	\end{equation}
	where $x_{k,i}$ denotes the $i$-th entry of $\mb x_k$ in (\ref{AoAEst})  and $\mb e_{k,u}^f$ denotes the corresponding subvector of  $\mb e_k$.  
	Letting the covariance matrices of the ``sources" $ {\bm \lambda}_{k }$ and ``errors" $\mb e_{k,u}^f$ be $\bm \Sigma_s$ and $\mb \Sigma_{e, u}^f$, respectively, we have 
	\begin{equation} \label{Rfu}
	\begin{aligned}
	\widetilde{\mathbf R}^{f,u}_{\bm \theta_R}&\triangleq  \mathrm{E} [\mb x_{k, u}^f \mb x_{k, u}^{f^H}] \\ & = \widetilde{\mb A} _R (\bm\theta_R)  \mb D^{(u-1)} \bm \Sigma_s   (\mb D^{(u-1)})^\ast  \widetilde{\mb A} _R^H (\bm\theta_R) + \mb \Sigma_{e, u}^f.  
	\end{aligned} 
	\end{equation}
	Define the forward covariance matrix as 
		\begin{equation} \label{Rf}
		\begin{aligned}
		\widetilde{\mathbf R}^{f}_{\bm \theta_R} & \triangleq \frac{1}{U} \sum_{u=1}^U \widetilde{\mathbf R}^{f,u}_{\bm \theta_R}  =\widetilde{\mb A} _R (\bm\theta_R) \bm\Sigma_s^f    \widetilde{\mb A} _R^H (\bm\theta_R) + \mb \Sigma_{e}^f   
		\end{aligned}
		\end{equation}
		with 
				\[ \bm\Sigma_s^f   \triangleq \frac{1}{U}\sum_{u=1}^{U}  \mb D^{(u-1)} \bm \Sigma_s   (\mb D^{(u-1)})^\ast,    \quad  	
 \mb \Sigma_{e}^f    \triangleq \frac{1}{U}\sum_{u=1}^{U}  \mb \Sigma_{e, u}^f. \] 
	In general, $\mb \Sigma_s^f$ has a higher rank than $\mb \Sigma_s$ when the ``sources" are correlated. This is beneficial for applying subspace methods for finding the angles.
	Similarly we can construct the $u$-th backward subarray as
	\begin{equation} 
	\begin{aligned}
	\label{xkb}
	\mb x_{k, u}^b & \triangleq \left[ x_{k, M-u+1}^*, x_{k, M-u}^*,  \cdots,  x_{k, M-u-\mathcal{S}}^* \right]^T  \\& =  \widetilde{\mb A} _R (\bm\theta_R)  \mb D^{-M+u}   {\bm \lambda}_{k}^\ast + \mb e_{k,u}^b. 
	\end{aligned}
	\end{equation}   
	Following the same assumption as for the forward subarray, we can verify   
	\begin{equation}  \label{Rbu}
	\begin{aligned}
	\widetilde{\mathbf R}^{b,u}_{\bm \theta_R} \triangleq & \mathrm{E} [\mb x_{k, u}^b \mb x_{k, u}^{b^H}]\\= &\widetilde{\mb A} _R (\bm\theta_R)  \mb D^{(-M+u)} \bm \Sigma_s^*   (\mb D^{(-M+u)})^\ast  \widetilde{\mb A} _R^H (\bm\theta_R) + \mb \Sigma_{e, u}^b   
	\end{aligned}
	\end{equation}
	where $ \mb \Sigma_{e, u}^b$ denotes the covariance matrix of $\mb e_{k,u}^b$.  
	 The backward covariance matrix can be defined similarly as  
		\begin{equation} \label{Rb}
		\begin{aligned}
		\widetilde{\mathbf R}^{b}_{\bm \theta_R} &\triangleq \frac{1}{U} \sum_{u=1}^U \widetilde{\mathbf R}^{b,u}_{\bm \theta_R}  = \widetilde{\mb A} _R (\bm\theta_R) \bm\Sigma_s^b    \widetilde{\mb A} _R^H (\bm\theta_R) + \mb \Sigma_{e}^b   
		\end{aligned}
		\end{equation}
		with \[\bm\Sigma_s^b = \frac{1}{U} \sum_{u=1}^{U}  \mb D^{(-M+u)} \bm \Sigma_s^\ast   (\mb D^{(-M+u)})^\ast, \quad \mb \Sigma_{e}^b    \triangleq \frac{1}{U}\sum_{u=1}^{U}  \mb \Sigma_{e, u}^b. \]
	Inspired by that (\ref{Rf}) and (\ref{Rb}) share the same signal subspace in the error-free case, the forward-backward spatially smoothed (FBSS) SCM can be used to estimate a subarray covariance matrix from the samples as  
	\begin{equation}  
	\begin{aligned}
	\widehat{\mathbf R}_{\bm \theta_R}^{SS} \triangleq \frac{1}{2KU} \sum_{k=1}^K \sum_{u=1}^{U} \left(\mb x_{k, u}^f \mb x_{k, u}^{f^H} + \mb x_{k, u}^b \mb x_{k, u}^{b^H} \right). 
	\end{aligned}
	\end{equation}
	 
Let $\mb J$ be the anti-diagonal identity matrix. Then the received signals of the $u$-th forward and $(U-u+1)$-th backward subarrays can be related as 
		\[ 
		\mb x_{k, U-u+1}^b  = \mb J (\mb x_{k, u}^f)^\ast,   \forall u,
		\]
		and
	\begin{align*}
\mb x_{k,  U-u+1}^b (\mb x_{k,  U-u+1}^b)^H  =& \mb J  (\mb x_{k, u}^f)^\ast  (\mb x_{k, u}^f)^T \mb J \\ =&\mb J  [(\mb x_{k, u}^f)  (\mb x_{k, u}^f)^H]^* \mb J, \forall u.
	\end{align*}
		This suggests that
the FBSS covariance matrix can  be alternatively obtained from  (\ref{RAhatL}) as  
\begin{equation}
\label{FFBSS} 
\begin{aligned}
 \widehat{\mathbf R}_{\bm \theta_R}^{SS}  = &    \mathrm{FBSS} ( \widehat{\mathbf R}_{\bm \theta_R} )   \\ 
\triangleq &  \frac{1}{2U}\sum_{ u=1}^{U} \left( \widehat{\mathbf R}_{\bm \theta_R}(u: u+\mathcal{S}-1, u:u+\mathcal{S}-1) \right.  \\& + \left. \mb J{\widehat{\mathbf R}_{\bm \theta_R}^{ {*}}} (u:u+\mathcal{S}-1,u:u+\mathcal{S}-1) \mb J \right).
\end{aligned}
\end{equation} 
After that, the root-MUSIC algorithm is applied on  $\widehat{\mathbf R}_{\bm \theta_R}^{SS}$ to estimate $\bm \theta_{R}$. 

In general, the resolution of angle estimation using root-MUSIC improves when the array aperture is larger and the robustness improves when there are more samples. With the above FBSS, root-MUSIC uses arrays consisting of  $\mathcal{S}=M-U+1$ antenna elements instead of the original $M$ antenna elements, but the number of samples is increased from $K$ to $2KU$. This sacrifices the array aperture and the resolution of angle estimation, but enhances the robustness so that the overall accuracy is generally improved, especially in large arrays. For estimating $\bm \phi_T$ using root-MUSIC, the construction of the required covariance matrix $\widehat{\mathbf R}_{\bm \phi_T}^{SS}$ follows the same way as $\widehat{\mathbf R}_{\bm \theta_R}^{SS}$ based on $\widehat{\bm {\mathcal H}}_0^H$ and the discussion is omitted for brevity.

\subsection{Stage 2: Estimation of IRS Angles and Composite Path Gains}

 \subsubsection{ {Training}} 
\label{irstraining}
We next estimate the IRS angles $\bm \psi_I$ and the composite path gains of the cascaded channel in (\ref{HParameterized}). The estimates $(\widehat{\bm \theta}_{R}, \widehat{\bm \phi}_{T})$ of the outer angles are used here to construct the hybrid precoder and combiner, respectively, for achieving beamforming gains. The desired precoder and combiner are given as 
\begin{equation} \label{PtWt} 
 \begin{aligned}
\widehat{\mb W} & = \mb A_R ( \widehat{\boldsymbol \theta}_R ) \in \mathbb{C}^{M \times L_G}, \\ 
\widehat{\mb P} &=\mb A_T( \widehat{\boldsymbol \phi}_T) \in \mathbb{C}^{K \times L_F}. 
\end{aligned}
\end{equation} 
They are implemented approximately as ${\mb P}$ and $\mb W$ using the fully connected hybrid transceivers and the PE-Altmin algorithm \cite{yu2016alternating}.   
 
Without loss of generality, we assume $L_G\le Q_r$, but the treatment can be extended for $L_G > Q_r$. 
There are in total $D$ steps in IRS angle training and each step spans $L_F$ channel uses. The precoder $\mb P$,  
combiner $\mb W$ and also the IRS phase shifts $\bm \Omega_d$ remain fixed for each step. The resulting effective channel  $\bm {\mathcal H}_{d}=\mb {G}\mb{\Omega}_{d}\mb F$.  During the $l$-th channel use of the $d$-th step, the received signal is given as  
\begin{align}
\mb y_{d,l}&=\mb W^H \bm {\mathcal H}_{d} \mb P \mb s_{{d},l}+\mb W^H\mb n_{{d},l}\in \mathbb{C}^{L_G \times 1},  
\end{align}
where $\mb s_{d,l} \in \mathbb C^{L_F \times 1}$ is the training symbol.
We choose $\{\mb s_{d, l}, l=1, 2, \cdots, L_F\}$ as columns of an $L_F \times L_F$ unitary matrix and without loss of generality, as the identity matrix. Furthermore, we assume $||\mb W||_F^2=L_G $, $||\mb P||_F^2=L_F $ and  $||\mb P\mb s_{d, l}||_F^2=1 $.  As a result, the observation at the receiver at the $d$-th step is equivalent to  
\begin{align}
\mb Y_{d} = [\mb y_{d,1}, \mb y_{d,2}, \cdots, \mb y_{d,L_F}] &= \mb W^H \bm {\mathcal H}_{d} \mb P  + \mb N_{d}' \in \mathbb{C}^{L_G \times L_F}. 
\end{align}
Let $\mb y_d = \mathrm{vec}(\mb Y_d)$. It can be verified that  
\begin{equation} 
\begin{aligned}
\mb y_d 
 &= \mathrm{vec} ( \mb W^H \bm {\mathcal H}_d \mb P)  + \mb n_d'  \\&= \left( (\mb F \mb P )^T \diamond (\mb W^H \mb G) \right)  \bm{\omega}_d  +   \mb n_d '   = \mb Z  \bm{\omega}_d +\mb n_d',  
\end{aligned}
\end{equation}    
where
\begin{equation} 
\label{Zmatrix}
\mb Z = \mb \Psi \bm \Gamma \mb  A_I^H(\bm \psi)  \in \mathbb C^{L_FL_G \times N},
\end{equation} 
\begin{equation} 
\label{mbPsi}
\mb \Psi =  \left(\mb P^T \mb A_T^*(\bm\phi_{T}) \right)  \otimes   \left(\mb W^H \mb A_R (\bm \theta_{R}) \right) \in \mathbb C^{L_FL_G \times L_F L_G}.   
\end{equation}

Recall that $\bm \Gamma$ in (\ref{Zmatrix}), {defined in (\ref{Gamma_Matrix})}, is diagonal. In the ideal case with infinite numbers of antennas, perfect outer angle estimation, and infinite resolution of the transceiver phase shifters, $\mb \Psi $ is an identity matrix. In this case, the IRS angles can be separately estimated from the corresponding rows of $\mb Z$, as in \cite{he2021channel}. However, $\mb \Psi$ is not identity or diagonal 
in practical systems. We therefore consider the joint estimation of the IRS angles based on $\mb Z$, which may improve performance.    

Variable phase shifts $\{\mb \Omega_d\}$ at the IRS are used during the $D$ training steps, yielding the overall observation    
\begin{equation} 
 \begin{aligned}
 \label{Y}
\mb Y&\triangleq [\mb y_1, \mb y_2, \dots, \mb y_D]  =\mb Z \bar{\mb \Omega}+\mb N' \in \mathbb{C}^{ L_G L_F \times D},
\end{aligned}
\end{equation} 
where $\bar{\mb \Omega}=[\bm \omega_1, \bm \omega_2, \dots, \bm\omega_D]\in \mathbb{C}^{ N \times D}$. If $D\ge N$, we can apply the LS method to estimate $\mb Z$, 
based on which the IRS angles $\bm \psi_I$ can be estimated using subspace methods. However, this may require substantial training when $N$ is large. 
In order to alleviate the training overhead, we choose $\bar{\mb \Omega}$ as 
\begin{align}\label{Omegabar}
\bar{\bm \Omega}=\begin{bmatrix}
\bm \Theta \\
\mb 0
\end{bmatrix} \in \mathbb{C}^{N\times D}, 
\end{align}
where $\bm \Theta \in \mathbb{C}^{D\times D}$ is a DFT matrix {with entries of unit magnitude} 
and $\mb 0$ denotes an all-zero matrix. This is equivalent to turning off $N-D$ elements of the IRS during the IRS angle training and sampling only $D$ elements of the IRS. For simplicity, we set the ``active" or ``switched-on" IRS elements to form a smaller ULA of size $D$. This reduces the aperture of the IRS and sacrifices the spatial resolution but may provide an economic way for training.  
 {A similar training design is used in \cite{chung2021atomic} to reduce the training overhead. }

\subsubsection{ Estimation of IRS angles}

In order to estimate $\bm \psi_I$ we first {apply the LS estimator to} obtain 
\begin{align} \label{Ztilde}
\widetilde{\mb Z} =\frac{1}{D} \mb Y \bm \Theta^H = \mb \Psi\mb \Gamma \widetilde{\mb A}^H_I (\bm \psi_I)+\widetilde{\mb N} \in \mathbb{C}^{L_{F}L_{G} \times D}, 
\end{align}
where  $ \widetilde{\mb A}_I (\bm \psi_I) \in \mathbb C^{D \times L_F L_G}$ contains the steering matrix for the sub-array of the IRS corresponding to its $D$ switched-on  elements and $\widetilde{\mb N}= \frac{1}{D} \mb N' \bm \Theta^H $. Now the IRS angles can be estimated based on  
\begin{align}\label{RPSI}
\widehat{\mb R}_{\bm \psi_I}=\frac{1}{L_{F}L_G} \widetilde{\mb Z}^H \widetilde{\mb Z}.   
\end{align}
Similarly to the outer angle estimation in Stage 1, in order to alleviate the issue of coherence of the ``source", FBSS can be applied to produce $\widehat{\mb R}_{\bm\psi_I}^{\mathrm{SS}} = \mathrm{FBSS} (\widehat{\mb R}_{\bm \psi_I})$ before applying root-MUSIC to produce the IRS angle estimate $\widehat{\bm \psi}_I$, where $\mathrm{FBSS(\cdot)}$ follows (\ref{FFBSS}). 
 
From (\ref{Y})-(\ref{RPSI}), it can be seen that the IRS angles can be estimated with a low computational complexity by integrating LS, FBSS and root-MUSIC. The training overhead is also reduced by switching on only a subarray at the IRS. Furthermore, as will be shown in the simulation results, the joint estimation of the IRS angles in the proposed solution is robust against the errors in the outer angle estimation and the finite size of the antenna arrays. It can thus improve the accuracy for cases with multiple paths as compared to the alternative treatment \cite{he2021channel} that estimates the IRS angles separately.

\subsubsection{ Estimation of composite path gains}

Once 
$ {\bm \theta}_{R}$, 
$  {\bm \phi}_{T}$, and
$ {\bm \psi}_I$ are estimated, the composite path gains $\mb \Gamma$ can be estimated by fitting
the received training signal to the model 
\begin{equation} 
\begin{aligned}   
  \mb y & \triangleq   \mathrm{vec}( \mb Y) = \bm \Phi  \bm  \gamma + \mb n'            
\end{aligned}
\end{equation} 
using LS, 
where 
 \begin{equation} 
\label{Phigamma}
\begin{aligned}
\bm \Phi 
 & =\left( \bar{\mb \Omega}^T  \mb A_I^\ast(  {\bm \psi}_I ) \right) \otimes  \left(\mb P^T \mb A_T^*( {\bm \phi}_{T}) \right)  \otimes   \left(\mb W^H \mb A_R (  {\bm \theta}_{R}) \right)   \\
\bm \gamma & \triangleq  \mathrm{vec}(\bm \Gamma).  
\end{aligned}
\end{equation}
Note that $\mb W$, $\mb P$ and $\bar{\bm \Omega}$ are known from IRS angle training, and the estimates of $\bm \psi_I$, $\bm \phi_T$ and $\bm \theta_R$ are also available. However, we have estimated the path angles separately for low complexity and thus they are not associated, which leads to an unknown sparsity pattern of $\bm \gamma$. For associating the angles and also for obtaining the composite path gains, we resort to a compressive sensing (CS) approach.  
Since $\bm \gamma \in \mathbb C^{L_F^2 L_G^2 \times 1}$ is a sparse vector and has only $L_GL_F$ non-zeros entries, the above problem can be solved efficiently using OMP as
\begin{align}
 	\widehat{\bm \gamma} = \arg \min_{\bm \gamma} ||\mb y- \widehat{\bm \Phi}  \bm \gamma ||^2, \qquad ||\bm \gamma ||_{0} =  L_GL_F.   
 	\label{gamma}  
\end{align}
where the ``dictionary" $\widehat{\bm \Phi} $ is obtained by replacing the angles in (\ref{Phigamma}) by their estimates.  
Once the gains are estimated, we can obtain the estimate of the cascaded channel matrix (\ref{HParameterized}) as 
\begin{equation}  
\label{HParameterizedEstULA}
\begin{aligned}
\widehat{\mb H}  =  ( \mb A_T^*( \widehat{\bm \phi}_{T})  \otimes \mb A_R (\widehat{\bm \theta}_{R}))  \mathrm{Mat}( \widehat{\bm \gamma}) \mb  A_I^H(\widehat{\bm \psi}_I).  
\end{aligned}
\end{equation}  
 Note that the estimated $ \mathrm{Mat}( \widehat{\bm \gamma})$ is not necessarily diagonal due to random permutation. The overall channel estimation process for the ULA case discussed in Section III.A-C is summarized in Algorithm 1.

\begin{algorithm}[t]
	\SetAlgoLined
	\caption{Cascaded channel estimation for IRS-aided hybrid MIMO systems employing ULAs at the transmitter, IRS, and receiver.}
	\SetKwInput{KwInput}{Input}                
	\SetKwInput{KwOutput}{Output}             
	\DontPrintSemicolon
	\textbf{Stage 1}:\\
	\KwInput{$\mb Y_0 ,\mb X_T$, and $ \mb X_R. $} 
	1.	Find $\widehat{\bm {\mathcal C}}_0^\star$ by solving (\ref{CstarHat}) using the GCG-ALTMIN algorithm \cite[Algorithm 1]{hu2018matrix}.\\
	2. Obtain the low-rank estimate $\widehat{\bm {\mathcal H}}_0$ using (\ref{Hl}).\\
	3. {Construct $\widehat{\mb R}_{{\bm \theta}_R}^{\mathrm{SS}}$ from $\widehat{\bm {\mathcal H}}_0$  using (\ref{RAhatL}) and (\ref{FFBSS}).}\\
	4. {Find $\widehat{\bm \theta}_R$ from $\widehat{\mb R}_{{\bm \theta}_{R}}^{\mathrm{SS}}$  using root-MUSIC.}\\
	5. {Construct $\widehat{\mb R}_{{\bm \phi}_T}^{\mathrm{SS}}$ from $\widehat{\bm {\mathcal H}}_0^H$ similar to (\ref{RAhatL}) and (\ref{FFBSS}).}\\
	6. {Find $\widehat{\bm \phi}_T$ from $\widehat{\mb R}_{{\bm{\phi}}_T}^{SS}$ using root-MUSIC.}\\
	\KwOutput{	$\widehat {\bm{\phi}}_T,\widehat{\bm {\theta}}_R$. }
	\textbf	{Stage 2}: \\
	\KwInput{ $\mb Y, \mb W, \mb P$, and $\bar{\mb \Omega}$.} 
	7. {Obtain $\widetilde{\mb Z}$ using (\ref{Ztilde}).}\\
	8. {Construct $\widehat{\mb R}_{\psi_I}^{\mathrm{SS}}$  from $\widetilde{\mb Z}$  using (\ref{RPSI}) and FBSS.}\\
	9. {Find  $\widehat{\bm \psi}_I$ from $\widehat{\mb R}_{\bm \psi_I}^{\mathrm{SS}}$ using root-MUSIC.}\\
	10. {Find $\widehat{\bm \gamma}$ using (\ref{gamma})}.\\
	11. {Construct the cascaded channel $\widehat{\mathbf H}$ using (\ref{HParameterizedEstULA}).}\\
	\KwOutput{$\widehat{\bm \psi}_I$, $\widehat{ \bm \gamma}$ and $\widehat{\mb H}$. }	
\end{algorithm}

\subsection{ Extension to UPA at IRS} 
\label{ExtUPAIRS}

\begin{figure}
	\centering
	\includegraphics[width=3.5in]{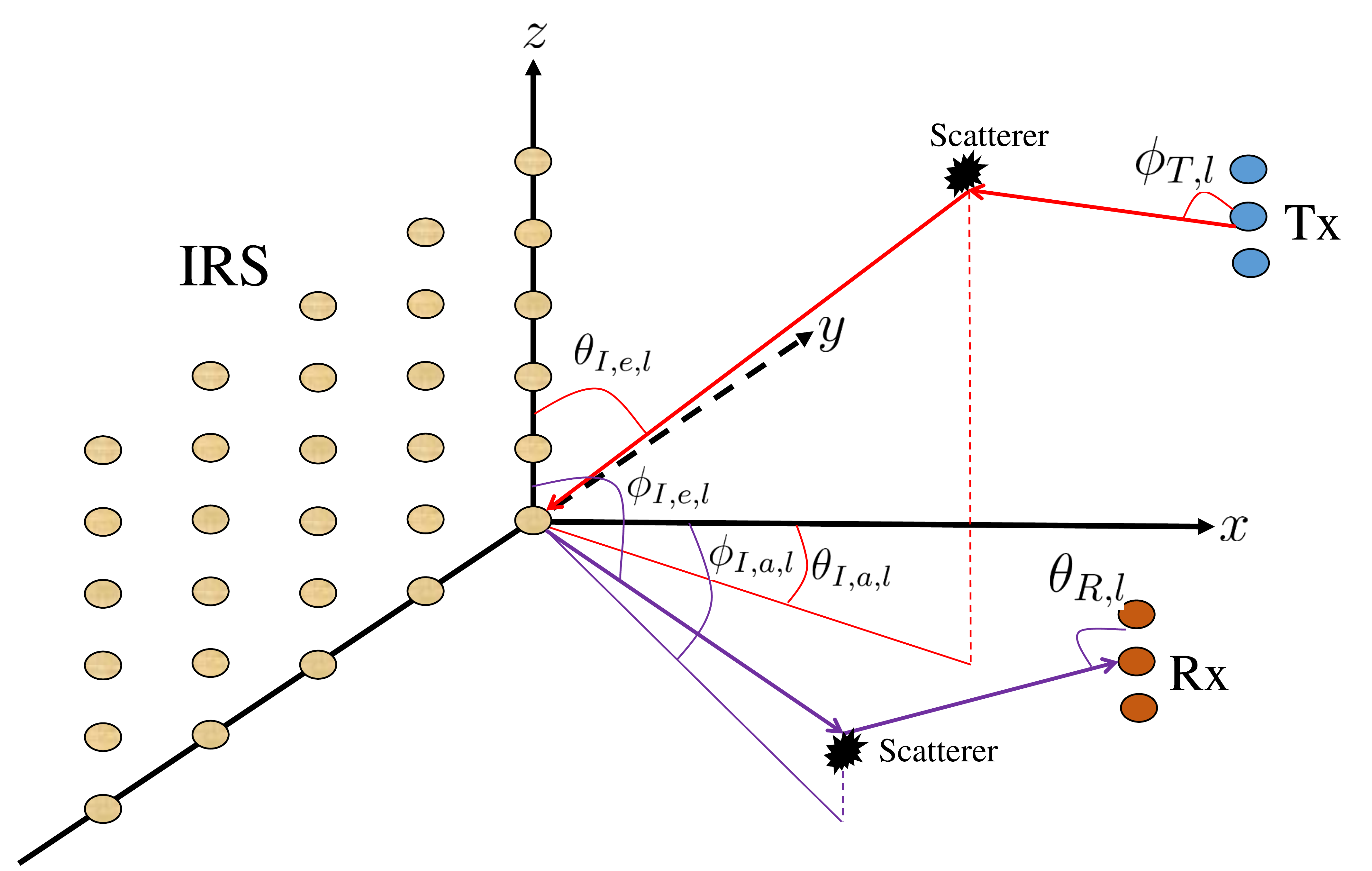}
	\caption{System model for the case with UPA at the IRS and ULAs at the transmitter and receiver. } 
	\label{sm}
\end{figure}
\begin{figure}
		\centering
		\includegraphics[width=2.5in]{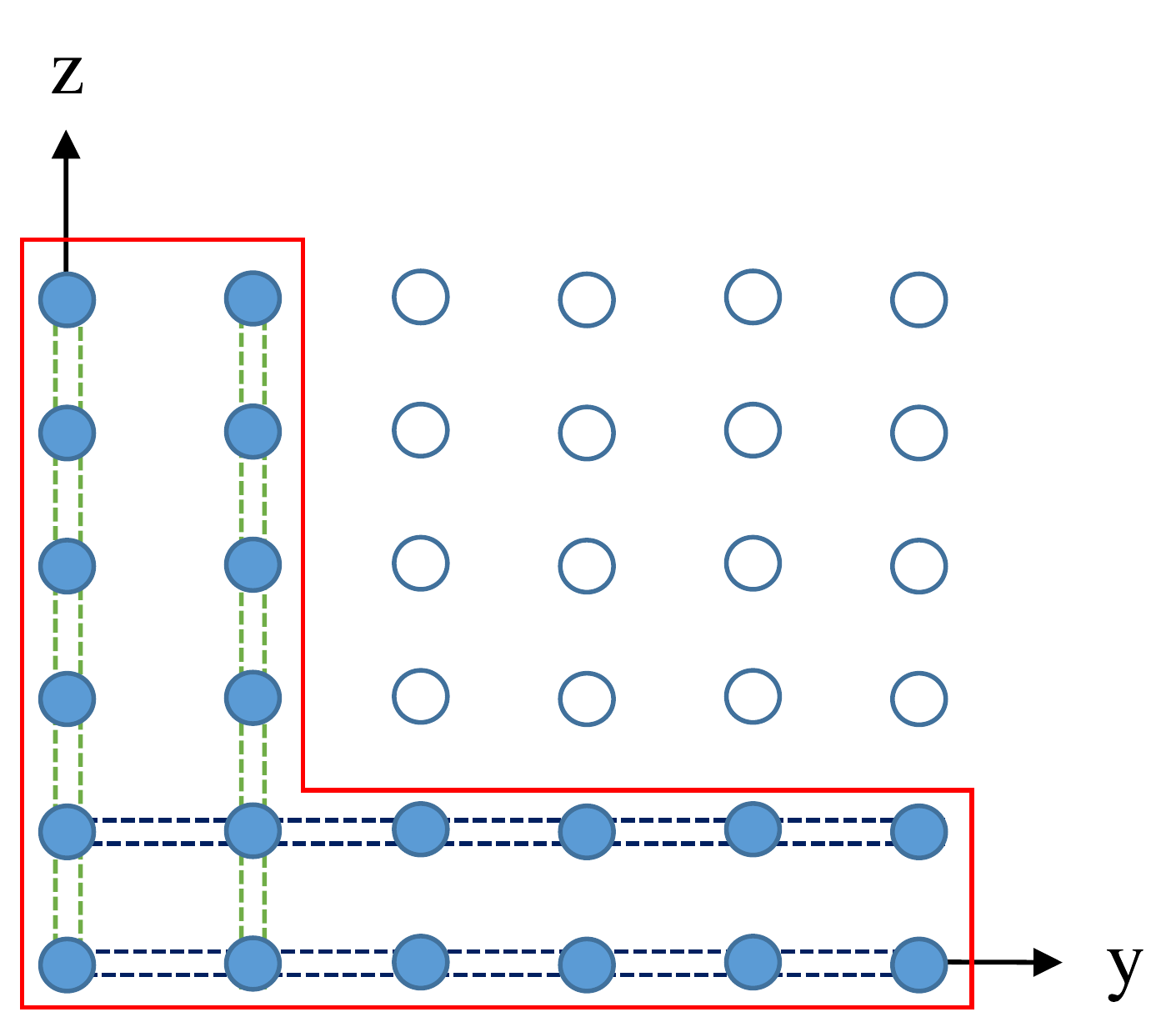}
		\caption{An example of IRS with an L-shaped subarray switched on, $N_y=N_z=6, J_y=J_z=2$, $N=36$ and $D=20$. }
		\label{LShaped}
	\end{figure}

 
In the above, we have assumed ULA at the transmitter, receiver, and IRS. We now extend the scheme to IRS equipped with UPA. When an $N_{y}\times N_{z}$ UPA is located on the $yz$ plane, as illustrated in Fig. \ref{sm}, the steering vector for a path with azimuth angle $\alpha_a$ and elevation angle $\alpha_e$ is given by 
\begin{align*}
\mb a(\alpha_a, \alpha_e)=&\mb a_y(\alpha_a,\alpha_e)\otimes\mb a_z(\alpha_e),  
\end{align*} 
where  
\begin{equation}
\begin{aligned}
\mb a_y(\alpha_a, \alpha_e)=&\frac{1}{\sqrt{N_{y}}}[1,e^{j\frac{2\pi}{\lambda_{c}}d\sin(\alpha_a)\sin(\alpha_e)}, \dots,  \\&
e^{j({N_{y}}-1)\frac{2\pi}{\lambda_{c}}d\sin(\alpha_a)\sin(\alpha_e)}]^T,   
\end{aligned}
\end{equation}
and
\begin{align}
\mb a_z(\alpha_e)=\frac{1}{\sqrt{N_{z}}}[1,e^{j\frac{2\pi}{\lambda_{c}}d\cos(\alpha_e)}, \dots
,e^{j({N_{z}}-1)\frac{2\pi}{\lambda_{c}}d\cos(\alpha_e)}]^T.
\end{align}
Let 
\begin{equation} \label{uv} 
u=\sin(\alpha_a)\sin(\alpha_e), \quad v=\cos(\alpha_e).
\end{equation} 
Then the steering vector  can be rewritten as 
\begin{align}
\mb a(u,v)=&\mb a_y(u)\otimes\mb a_z(v), 
\end{align} 
where 
\begin{equation}
 \begin{aligned}
\mb a_y(u)=&\frac{1}{\sqrt{N_{y}}}[1,e^{j\frac{2\pi}{\lambda_{c}}d u}, \dots
,e^{j({N_{y}}-1)\frac{2\pi}{\lambda_{c}}d u}]^T, \\
\mb a_z(v)=&\frac{1}{\sqrt{N_{z}}}[1,e^{j\frac{2\pi}{\lambda_{c}}dv}, \dots
,e^{j({N_{z}}-1)\frac{2\pi}{\lambda_{c}}dv}]^T. 
 \end{aligned}
\end{equation}
Denote  by $(\theta_{I, a, l}, \theta_{I, e, l})$ the pair of the azimuth and elevation components of the AoA for the $l$-th path impinging on the IRS and similarly for those of the AoD of the $l$-th path departing the IRS by $(\phi_{I, a, l}, \phi_{I, e, l})$. Following the change of variables as in (\ref{uv}), we can rewrite the steering matrix in (\ref{Zmatrix}) for the UPA as 
	\begin{equation}
\begin{aligned} \nonumber 
\mb A_I (\mb u,\mb v) =&\left(\mb A_I^H (\mb u_F,\mb v_F)\diamond \mb A_I^T(\mb u_G,\mb v_G)\right)^T\\=& [\mb a(u_{1,1},v_{1,1}), \dots, \mb a(u_{i,j},v_{i,j}),\dots,  \\& \mb a(u_{L_F,L_G},v_{L_F,L_G})],
\end{aligned}	
\end{equation}
where 
	\begin{equation}
	\begin{aligned}
	u_{i,j}=& \sin(\phi_{I, a, i})\sin( \phi_{I, e, i}) - \sin(\theta_{I, a, j})\sin( \theta_{I, e, j}), \\
	v_{i,j}=& \cos(\phi_{I, e, i}) -\cos(\theta_{I, e, j}).
	\end{aligned}
	\end{equation} 
In this case, the estimation of the outer angles follows that for the ULA case in Section III-B. The IRS angle estimation now amounts to estimating $(\mb u, \mb v)$. In order to achieve high accuracy with low training overhead and low computational complexity, we propose to switch on only an L-shaped subarray as illustrated in Fig. \ref{LShaped} during the IRS angle estimation. 
The corresponding IRS phase shift matrix $\bar{\bm \Omega}$ during the Stage-2 training can still be expressed as (\ref{Omegabar}). Specifically, the DFT matrix $\bm \Theta \in \mathbb{C}^{D\times D}$ is applied to the IRS elements in the L-shaped subarray only (which are indexed by $1\le n \le D$) while the zero matrix $\mb 0$ is applied to the remaining IRS elements. 
The observation after completing the Stage-2 training can then be modeled similarly to (\ref{Ztilde}) as  
\begin{align}
\widetilde{\mb Z}= \mb\Psi\mb \Gamma \widetilde{\mb A}^H_I(\mb u,\mb v)+\widetilde{\mb N}
\label{Z}  \in \mathbb C^{L_F L_G \times D }, 
\end{align}
where $\widetilde{\mb A}_I(\mb u,\mb v) \in \mathbb C^{D \times L_F L_G}$ is the array response matrix of the L-shaped subarray. 
It is clear that the IRS angle information is embedded in the row subspace of $\widetilde{\mb Z}$.  

Note that the L-shaped subarray consists of partially overlapped ULAs along the $y$ and $z$ axes. Denote by $\mathcal{I}_{y, j}$ the indices of the IRS elements of the $j$-th ULA parallel to the $y$-axis in Fig. \ref{LShaped}. We choose the observations in $\widetilde{\mb Z}$ corresponding to the $J_y$ ULAs and  stack them as 
\begin{align}
\widetilde{\mb Z}_y= \begin{bmatrix} \widetilde{\mb Z} (:, \mathcal{I}_{y, 1}) \\
\widetilde{\mb Z}(:, \mathcal{I}_{y, 2})\\
\vdots\\
\widetilde{\mb Z}(:,\mathcal{I}_{y, J_y})
\end{bmatrix} \in \mathbb{C}^{J_y L_FL_G  \times N_y}. 
\label{Zu}
\end{align}   
In order to estimate $\mb u$, we first compute 
	\begin{align}
	\mb R_u=\frac{1}{J_yL_FL_G}\widetilde{\mb Z}_y^H\widetilde{\mb Z}_y, 
	\end{align}
then obtain $\mathrm{FBSS(\mb R_u)}$ and finally apply root-MUSIC. We can apply a similar procedure to find $\mb v$ by using the observations corresponding to the ULAs parallel to the $z$ axis at the IRS. The adopted L-shaped subarray, which has been examined in \cite{xi2014computationally,wei2014pair,gu2015joint}, achieves larger spatial aperture along the $y$ and $z$ axis. This results in higher resolution in the estimation of the azimuth and elevation angles as compared to rectangular subarrays with the same number of elements.

Unlike joint 2D spectrum estimation \cite{haardt19952d}, we here estimate $\mb u$ and $\mb v$ separately for reduced computational complexity. This, however, does not give associated estimates of $\mb u$ and $\mb v$. Again the estimation of the composite path gains and association of the estimates of the path angles can be achieved by solving the CS problem in (\ref{gamma}), with the dictionary updated for the UPA case  as
 \begin{equation} \label{PhiIRS}
\begin{aligned}
\widehat{\bm \Phi} & =\left( \bar{\mb \Omega}^T\mb A_I^\ast( \widehat{\mb u}, \widehat{\mb v} ) \right) \otimes  \left(\mb P^T \mb A_T^*( \widehat{\bm\phi}_{T}) \right)  \otimes   \left(\mb W^H \mb A_R ( \widehat{\bm \theta}_{R}) \right).
\end{aligned}
\end{equation}
Finally the cascaded channel is  reconstructed as  
\begin{equation}  
\label{HParameterizedEst}
\begin{aligned}
\widehat{\mb H}  =  ( \mb A_T^*( \widehat{\bm \phi}_{T})  \otimes \mb A_R (\widehat{\bm \theta}_{R}))  \mathrm{Mat}( \widehat{\bm \gamma}) \mb  A_I^H(\widehat{\mb u},\widehat{\mb v}).  
\end{aligned}
\end{equation} 
Treatments same as this subsection can be applied to the case with UPA at the transmitter and/or receiver. The details are omitted for conciseness.

\subsection{Computational Complexity} 
 
The computational cost of the proposed estimator is kept low. 
For Stage 1 (outer angle estimation), the GCG-ALTMIN algorithm has a complexity of {$\mathcal{O}(I L_m^3KM+ L_m^2(K+M)+L_mKM)$ where $L_m=\min(L_F, L_G)$ and } $I$ is the number of iterations of the alternate minimization step of GCG-ALTMIN, which is low when the channel is low-rank with small $L_F$ and $L_G$. In large systems, the overall complexity is dominated by the root-MUSIC algorithm with complexity $\mathcal{O}(M^3+K^3)$. Since $\bar{\mb \Omega}$ is constructed using the DFT matrix, (\ref{Ztilde})  has a low cost. 

In Stage 2, the inner angle estimation with ULA at the IRS  using root-MUSIC has a complexity of $\mathcal{O}(D^3)$. The path gains are estimated using the OMP algorithm with a complexity of $\mathcal{O}(L_F^3 L_G^3D)$ which is also low when the channel has a small number of paths. 
For the case with UPA at the IRS, the proposed estimator has a complexity of $\mathcal{O}(N_v^3+N_u^3)$ for estimating the IRS angles in Stage 2. The composite path gain estimation involves a larger dictionary for the OMP algorithm due to the increased number of angles, but still requires the same order of complexity $\mathcal{O}(L_F^3  L_G^3 D)$.

\subsection{Training Overhead} 
The proposed channel estimation scheme has a training overhead of  
\begin{align}
\label{Toverhead}
T =S+DL_F  
\end{align}
channel uses, with $S$ channel uses for estimating the outer angles in Stage 1 and  $D L_F$ channel uses for estimating the IRS angles in Stage 2. In general, $S$ scales with the rank of the effective channel matrix $\bm {\mathcal  H}_0$ in (\ref{mathcalH0}) for a given accuracy of estimating $\bm {\mathcal  H}_0$ using IMC. Meanwhile, a larger $S$ results in better estimation of $\bm {\mathcal  H}_0$ and the outer angles $\widehat{\bm \theta}_R$ and $\widehat{\bm \phi}_T$. The overhead of Stage 2 scales linearly with the number of paths $L_F$ of the TX-IRS channel $\mb F$ and the number $D$ of the IRS elements switched on. Given the channel, a larger $D$ generally leads to improved estimation of the inner angles, i.e., the angle differences at the IRS. Based on the above analysis, the training overhead of the proposed solution can be kept low when the channel has a small number of paths, as typically seen in mmWave and THz systems.

\section{Simulation Results}

This section presents the simulation results. For the case with a ULA at the IRS, the path gains $\gamma_{F,l}$ and $\gamma_{G,l}$ in (\ref{Fmat}) and (\ref{Gmat}) follow $\mathcal{CN}(0,1)$ while all the AoAs and AoDs are uniformly distributed in $[30^\circ,150^\circ]$. For the case with a UPA at the IRS, the azimuth angles and elevation angles are uniformly distributed in $[-90^\circ,90^\circ]$ and $[30^\circ,150^\circ]$, respectively. Define the pilot-to-noise-ratio as
 $(\mathrm{PNR})=10\log_{10}(\frac{1}{\sigma_n^2})$. The normalized mean squared error (NMSE) for estimating the cascaded channel is evaluated as the average of  $\frac{||{\mb H} -\widehat{\mb H} ||_F^2}{||{\mb H} ||_F^2}$. 
The mean squared error (MSE) for angles $\bm \theta_{R}$, $\bm \phi_{T}$, $\bm \psi_I$ and gains $\bm \gamma$ are estimated, respectively, by averaging $\frac{||\cos(\bm \theta_{R})  -\cos(\widehat{\bm \theta}_{R}) ||_F^2}{L_G}$, $\frac{||\cos(\bm \phi_{T}) -\cos(\widehat{\bm \phi}_{T}) ||_F^2}{L_F}$, $\frac{||\cos(\bm \psi_I)  -\cos(\widehat{\bm \psi}_I) ||_F^2}{L_{F}L_G} $, and  
$\frac{||\bm \gamma  -\widehat{\bm \gamma} ||_F^2}{L_{F}L_G}$. For the UPA case, $\frac{||\mb u  -\widehat{\mb u} ||_F^2}{L_{F}L_G}$ and $\frac{||\mb v  -\widehat{\mb v} ||_F^2}{L_{F}L_G}$ are averaged to measure the MSE of estimating $\mb u$ and $\mb v$, respectively.

\subsection{ULA at the IRS} 
We first consider the case with ULAs at the transmitter, receiver and IRS. We compare the proposed estimator with the following two alternatives: 

\begin{itemize} 
\item  \emph{The ANM-based two-stage estimator of \cite{he2021channel}}: At stage 1, randomly generated precoding and combining matrices $ \widetilde{\mb P}$ and $\widetilde{\mb W} $ are applied to produce the training observation $\widetilde{\mb Y}_0 =  \widetilde{\mb W}^H \mb G \mb{\Omega}_0\mb F  \widetilde{\mb P}+ \mb N \in \mathbb C^{\frac{ S Q_r }{K} \times K}$ for a fixed, randomly generated $\mb \Omega = \mathrm{diag}(\bm \omega_0)$, followed by ANM for estimating $\bm \theta_R$ and $\bm \phi_T$. In order to compare under the same training overhead, we set $ \widetilde{\mb P} \in \mathbb C^{K \times K}$ and $ \widetilde{\mb W} \in \mathbb C^{M \times \frac{ S Q_r }{K}}$ at Stage 1. The estimation accuracy of $\bm\phi_{T}$  and $\bm\theta_{R}$ depends on the numbers of precoder and combiner vectors, respectively.  At Stage 2, the precoder and combiner are redesigned using $\widehat{\bm \theta}_R$ and $\widehat{\bm \phi}_T$ in the same way as in Section \ref{irstraining}. However, the IRS phase shifts vary randomly at $D$ steps with all the IRS elements switched on. ANM is applied to estimate each IRS angle separately. This ANM-based estimator has a complexity of   $\mathcal{O}(( \max\{K+SQ_r/K,  M+K,  N+1\})^{3.5} )$ when semidefinite programming (SDP) is applied to solve the ANM problems. This is generally more complex than the proposed estimator, especially for large systems. 
   
\item \emph{The LS estimator}: Approximately unitary precoders $\widetilde{\mb P}  \in \mathbb{C}^{K\times K}$ and combiners $ \widetilde{\mb W} \in \mathbb{C}^{M\times M}$ are implemented using the PE-Altmin algorithm for the hybrid transmitter and receiver, and the IRS phase shifts are selected as the columns of the DFT matrix $ \bar{\bm \Omega} \in \mathbb{C}^{N\times N}$. The training overhead is $T_{\mathrm{LS}}=KNM/Q_r$ channel uses.  

\end{itemize} 

\begin{figure} 
	\centering
	\begin{subfigure}[b]{0.5\textwidth}
		\centering
		\includegraphics[width=3.5in]{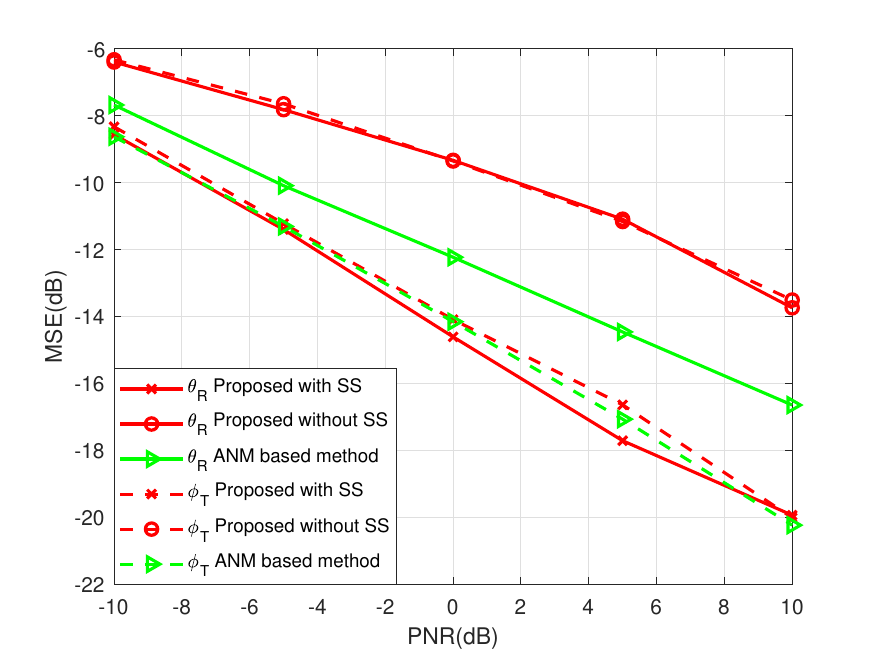}
		\caption{MSE of  outer angle estimation versus $\mathrm{PNR}$}
	\end{subfigure}
	
	\begin{subfigure}[b]{0.5\textwidth}
		\centering
		\includegraphics[width=3.5in]{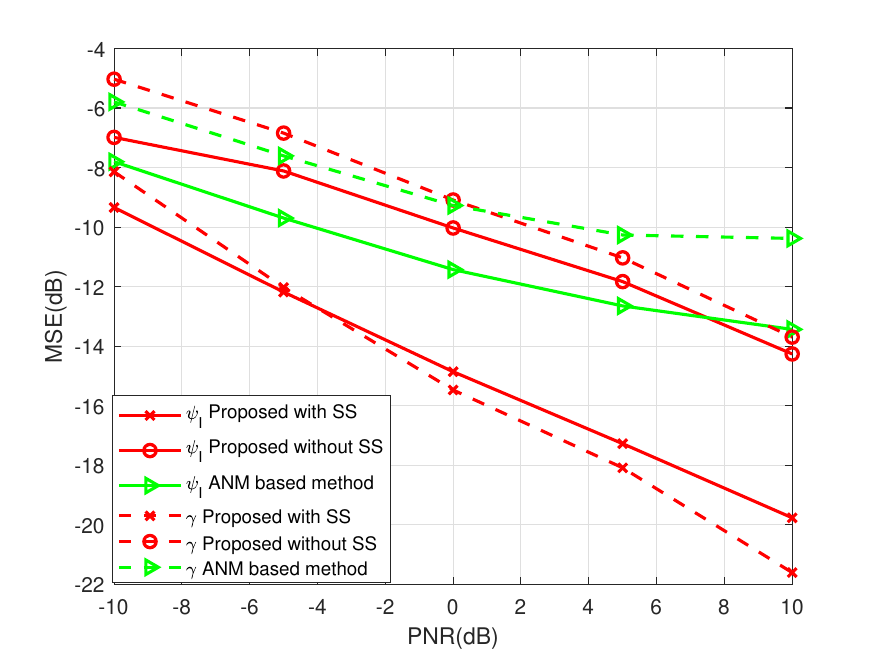}
		\caption{MSE of  IRS angle and composite path gain estimation versus $\mathrm{PNR}$}
	\end{subfigure}
	\begin{subfigure}[b]{0.5\textwidth}
		\centering
		\includegraphics[width=3.5in]{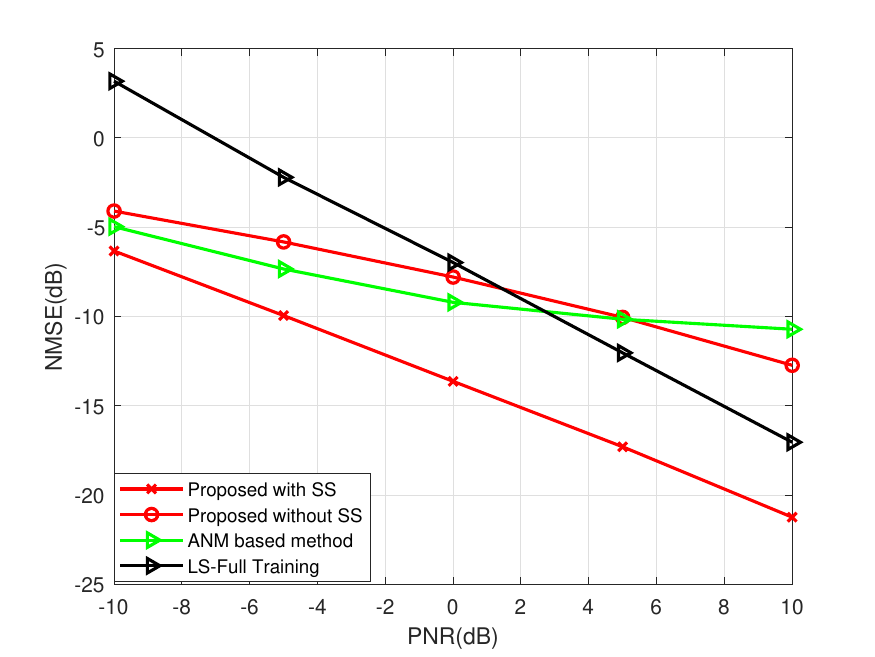}
		\caption{NMSE of cascaded channel matrix estimation versus $\mathrm{PNR}$}
	\end{subfigure}
	\caption{Performance of the path angle, gain and cascaded channel estimation with $L_F=2, L_G=2$, ULAs at the transmitter, receiver and IRS,  $K=M=16, N=32, Q_r=Q_t=2$. The proposed and the ANM-based methods apply the two stage training with $S=64, D=16$ and $T=S + D L_F = 96$ while the LS estimator uses a much higher training overhead of $T_{\mathrm{LS}}=4096$.}
	\label{ULA2path}
\end{figure}

Fig. \ref{ULA2path} compares the performance for $L_F=L_G=2$. It can be seen that the proposed estimator achieves good accuracies for estimating the channel parameter and the overall cascaded channel. FBSS effectively improves the angle estimation. The proposed estimator achieves significantly more accurate estimation of the inner angles $\bm \psi_I$ and path gains as compared with the ANM-based estimator. This is because the latter estimates the inner angles separately and employs the assumption that both $\mb A^H(\widehat {\bm \phi}_{F}) \mb P $ and $\mb W^H \mb A(\widehat{\bm \theta}_{G})$ as identity matrices. This assumption is valid for single-path scenarios where $L_F=L_G=1$, $\mb A^H(\widehat {\bm \phi}_{F}) \mb P $ and $\mb W^H \mb A(\widehat{\bm \theta}_{G})$ are close to unity. When there are multiple paths, i.e., $L_F > 1$ or $L_G > 1$, the identity matrix assumption requires perfect outer angle estimation and infinite sizes for the transmitter and receiver arrays. Otherwise, the data used for estimating a path also include the leakage from other paths, which leads to suboptimal parameter estimation suffering from interference between paths. The proposed estimator jointly estimates the inner angles and thus it does not rely on such an assumption and achieves more robust performance. This also translates into the improvement of the performance of the cascaded channel estimation. 
 Both the proposed and the ANM-based estimators significantly outperform the LS estimator when the PNR is low. The latter requires a much higher training overhead but does not benefit from the channel sparsity and the knowledge of the array responses. 
  
The channel estimation performance versus training overhead $T$ is demonstrated in Fig. \ref{Toverhead}, where $T$ is varied by varying $D$ in Stage 2. The results suggest that the performance of the proposed method generally improves when $T$ increases. When FBSS is used, around $4 \mathrm{dB}$ gain can be achieved for the proposed solution when $T$ varies from $88$ to $128$. The ANM method shows a more stable performance with respect to $T$ because the separate estimation of the inner angles is influenced by the leakage between the paths which can not be effectively mitigated by increasing the training data. 

 	\begin{figure} 
 		\centering
 		\includegraphics[width=3.5in]{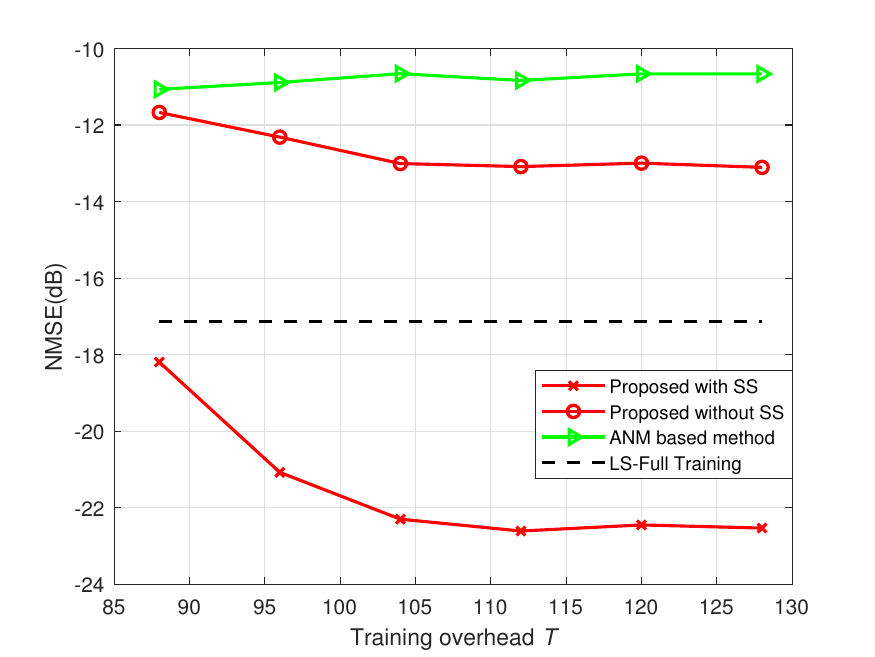}
 		\caption{Channel estimation performance versus training overhead $T=S+L_FD$ at $\mathrm{PNR}=10\ \mathrm{dB}$ with $L_F= L_G=2$, ULAs at the transmitter, receiver and IRS,  $K=M=16, N=32$, and $Q_r=Q_t=2$. The proposed method and the ANM-based method apply the two-stage training with $S=64$ fixed for Stage 1 and $D$ varying from $12$ to $32$ for Stage 2, while the LS estimator has a fixed training overhead of $T_{\mathrm{LS}}=4096$.}
 		\label{Toverhead}
 	\end{figure}

\begin{figure} 
	\centering
	\includegraphics[width=3.5in]{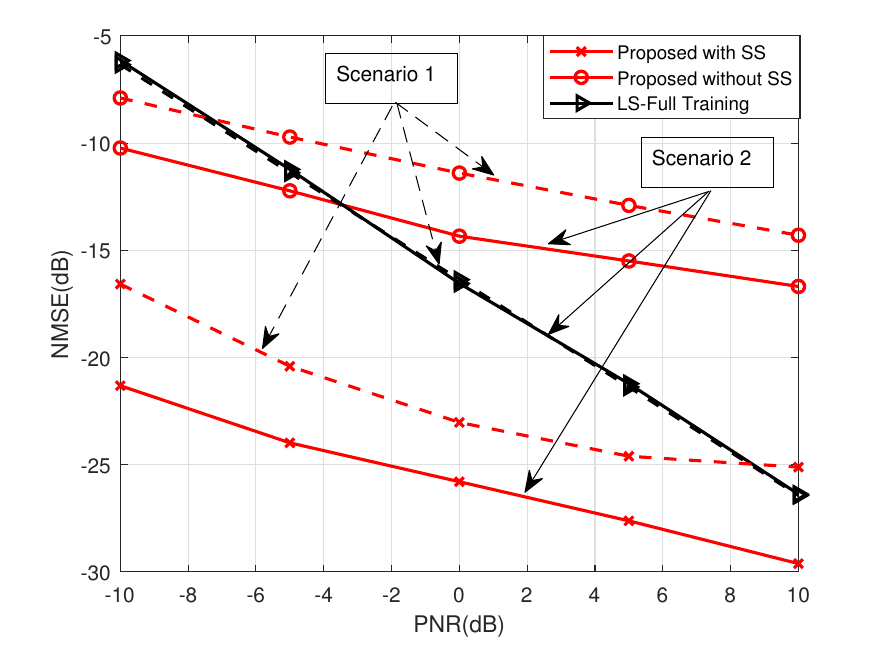}
	\caption{Channel estimation performance versus $\mathrm{PNR}$ with $L_F= L_G=4$. Scenario 1: $K=M=32, N=128, Q_r=Q_t=4$, $S=128, D=72$, and $T=S + D L_F =416$,  and $T_{\mathrm{LS}}=32768$. Scenario 2: $K=M=64, N=128, Q_r=Q_t=8$, $S=256, D=64$, and $T=S + D L_F =512$, and $T_{\mathrm{LS}}=65536$. }
	\label{4x4}
\end{figure}

The channel estimation performance for an example with $L_F=L_G=4$ is presented in Fig \ref{4x4}. There are two scenarios considered, where the proposed estimator performs better for Scenario 2 than Scenario 1 due to larger  $K$ and $M$. With far less training overhead, the proposed estimators with FBSS outperform the LS-based solution for both scenarios. The results suggest that the proposed solutions are effective when the TX-IRS and IRS-RX channels are low-rank (i.e. $L_F \ll \min\{K, N\}$ and $L_G \ll \min\{M, N\}$). This makes the proposed estimator a viable candidate for mmWave and THz channels where the number of channel paths is  considerably less than the channel dimensions. The ANM-based solution is skipped in the comparison due to the prohibitive simulation time requirements for the scenarios considered in Fig. \ref{4x4}.
 
\begin{figure} 
	\centering
	\includegraphics[width=3.5in]{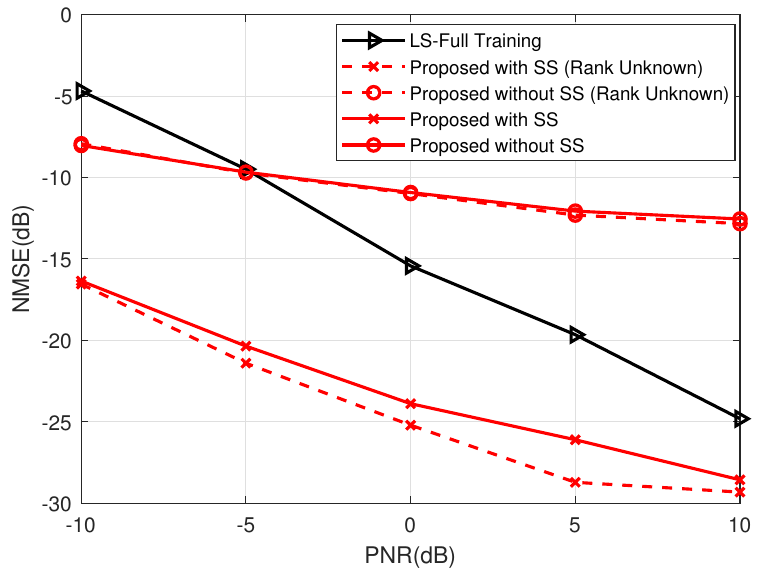}
	\caption{Average channel estimation performance versus $\mathrm{PNR}$ with $K=M=32, N=128, Q_r=Q_t=4$, $S=128, D=64$, and $T_{\mathrm{LS}}=32768$. The actual numbers of paths $L_F$ and $L_G$ are uniformly distributed in $\{2,3, 4\}$. When the actual rank $L_F=2,3$ and $4$ is perfectly known, the overall training overhead $T=256$, $320$, to $384$, respectively, and when $L_F$ is assumed to be $4$, we have $T=384$.}
	\label{Rank_test}
\end{figure}

In the above, we have assumed that the numbers of paths $L_F$ and $L_G$ are known as \emph{a priori} information, similarly to \cite{guo2017millimeter, he2021channel, schroeder2022two}. Such \emph{a priori} information may be set based on typical values of the channel models at given frequencies and settings (indoor, outdoor, etc.) \cite{gao2015mmwave}, but their perfect acquisition is still an open problem.  
However, the proposed solution can perform well when large enough values of $L_F$ and $L_G$ are assumed. In order to demonstrate this, let us consider a scenario where 
the actual values of $L_F$ and $L_G$ are random and uniformly distributed in $\{2,3, 4\}$, but they are assumed  to be fixed at $L_F=L_G=4$. In Fig. \ref{Rank_test}, we compare the resulting channel estimation performance with the case where the values of $L_F$ and $L_G$ are perfectly known. Note that the overall training overhead $T=S+DL_F$ increases linearly with the assumed value of $L_F$. 
The simulation results suggest that the proposed estimator can work well even with an overestimated value of $L_F$. 
The increased training overhead with the overestimated value of $L_F$ also leads to a slight performance gain.

\begin{figure} 
	\centering
	\begin{subfigure}[b]{0.5\textwidth}
		\centering
		\includegraphics[width=3.5in]{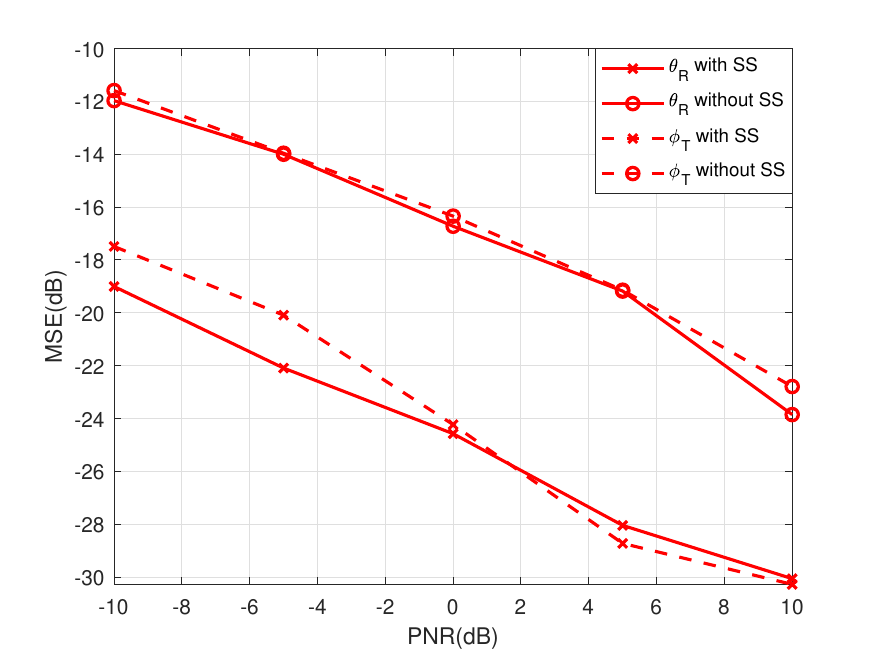}
		\caption{MSE of outer angle estimation versus PNR}
	\end{subfigure}
	
	\begin{subfigure}[b]{0.5\textwidth}
		\centering
		\includegraphics[width=3.5in]{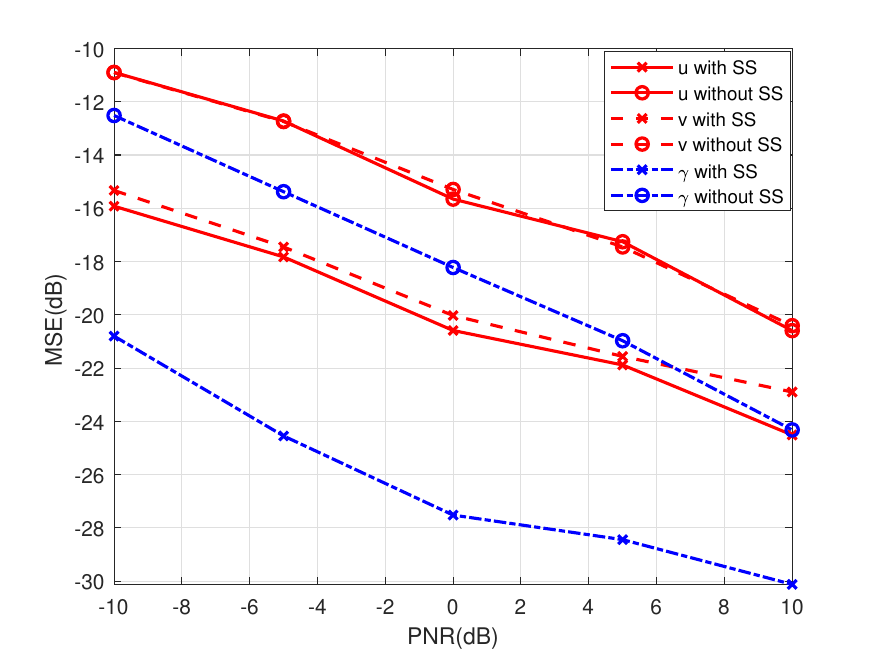}
		\caption{MSE of IRS angle and composite path gain estimation versus PNR}
	\end{subfigure}
	\begin{subfigure}[b]{0.5\textwidth}
		\centering
		\includegraphics[width=3.5in]{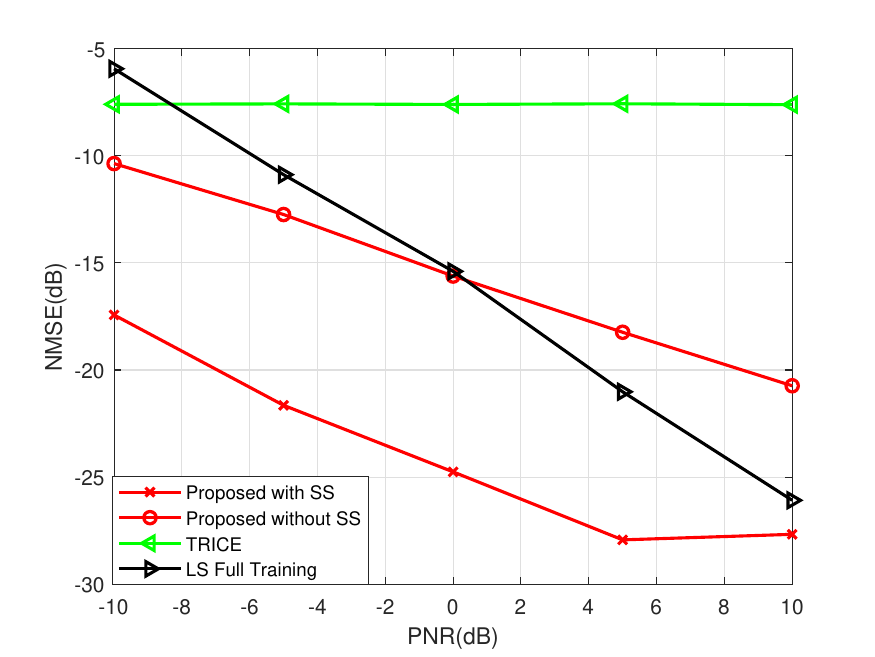}
		\caption{NMSE of cascaded channel matrix estimation versus PNR}
	\end{subfigure}
	\caption{Performance of the path angle, gain and cascaded channel estimation for a system with a UPA at the IRS, ULAs at the transmitter and receiver, and $L_F=2, L_G=2$. $K=M=32, N=256, Q_r=Q_t=4$. For the proposed estimator, $S=128, D=31, N_y=N_z=16, J_y=J_z=1$ and $T = S+ L_F D = 190$. For the TRICE estimator $K_0=M_0=16$, $N_0=32$, $T_{\mathrm{TRICE}}=2048, G_K=G_M=64, G_N=128 \times 128$. For the LS estimator, $T_{\mathrm{LS}}=65536$.} 
	\label{UPA1616}
\end{figure}

\subsection{UPA at the IRS} 

A case with UPA at the IRS is demonstrated in Fig. \ref{UPA1616} where a $16 \times 16$ UPA with $N=256$ is employed at the IRS. For comparison with the proposed estimator, we consider the LS estimator and the CS-based TRICE estimator of \cite{ardah2021trice}. Similarly to the proposed estimator, the TRICE estimator first estimates the outer angles followed by estimating the IRS angles and composite path gains using CS. However, it has a single stage of training and does not exploit the estimated outer angles while designing the training precoder and combiner. Due to the lower beamforming gains, the TRICE estimator requires a higher training overhead. Denote by $\bar{\mb P} \in \mathbb{C}^{K \times K_0}$, $\bar{\mb W } \in \mathbb{C}^{M \times M_0}$ and the IRS phase shift matrix $\bar{\bm \Omega} \in \mathbb{C}^{N \times N_0}$, where $K_0$, $M_0$ and $N_0$ represent the numbers of precoders, combining vectors and IRS states, respectively. They are used by the TRICE estimator to produce $K_0 M_0N_0$ observations using  $K_0 M_0N_0/Q_r$ channel uses. The TRICE estimator has a complexity of $\mathcal{O}(L_GL_F (K_0M_0 G_MG_T  +  N_0G_I))$, where $G_K$, $G_M$ and $G_N$ represent the grid sizes for the AoD, AoA and IRS angle, respectively. 
In our simulations,  $\bar{\mb P}$ and $\bar{\mb W}$ are taken from approximated unitary matrices (randomly constructed in the same way as $\mb X_T$ and $\mb X_R$ in (\ref{IMCprob})) and the IRS phase shift matrix $\bar{\bm \Omega} $ has phase shifts uniformly distributed over $[0,2\pi)$. From Fig. \ref{UPA1616}, the outer angle estimation with the proposed estimator has similar performance as for the case with ULA at the IRS. High accuracy for estimating $\mb u$ and $\mb v$ is also achieved, and the FBSS is effective for improving the performance with the L-shaped array adopted at the IRS. With FBSS, the proposed estimator achieves an overall performance of cascaded channel estimation significantly better than the alternative estimators, even at a much lower training overhead. This is due to the two-stage training which benefits from the beamforming gains at Stage 2, and the super-resolution estimation of the outer and IRS angles.  

\section{Conclusions}

In summary, we have presented a parametric method for estimating the cascaded channel for fully passive IRS-aided MIMO systems with hybrid transceivers. The proposed estimator benefits from the low-rank nature of the channel and the knowledge of the array responses. It provides a low-complexity, multiple-stage  solution using simple yet effective tools, including inductive matrix completion (IMC), forward-backward spatial smoothing (FBSS), and the root-MUSIC algorithm. This solution can progressively obtain the channel parameters and the training and estimation process adapts to the knowledge generated, which not only provides better estimation performance but also reduces the training overhead. As seen from the simulation results, the proposed estimator outperforms several recently studied solutions in estimation accuracy. The overall computational complexity of the proposed estimator is also kept low.

\section*{Acknowledgment}
The authors wish to thank the associate editor and the anonymous reviewers for their helpful comments. 

\bibliography{References}

\end{document}